\documentclass[10pt]{iopart}

\usepackage{iopams}
\usepackage[T1]{fontenc}
\usepackage[utf8]{inputenc}
\usepackage{graphicx}
\usepackage{tabularx}
\usepackage{color}
\usepackage{textcomp}
\usepackage{enumerate}
\usepackage{todonotes}
\usepackage[draft]{hyperref}

\renewcommand{\vec}[1]{\mathbf{#1}}

\providecommand{\xiaoran}[1]{#1}

 \providecommand{\RV}[1]{\textcolor{black}{#1}}

\begin{document}

\title[]{A computational study of positive streamers interacting with
  dielectrics}

\author{Xiaoran Li$^1$, Anbang Sun$^{1}$, Guanjun Zhang$^1$, Jannis Teunissen$^{2,3}$}

\address{$^1$\RV{State Key Laboratory of Electrical Insulation and Power Equipment, School of Electrical Engineering,
	Xi'an Jiaotong University, Xi'an, 710049, China}\\
  $^2$Centrum Wiskunde \& Informatica, Amsterdam, The Netherlands\\
  $^3$Centre for Mathematical Plasma-Astrophysics, KU Leuven, Belgium}
\ead{anbang.sun@xjtu.edu.cn,jannis@teunissen.net}
\vspace{10pt}
\begin{indented}
\item[]\today
\end{indented}

\begin{abstract}
  We use numerical simulations to study the dynamics of surface discharges,
  which are common in high-voltage engineering. We simulate positive streamer
  discharges that propagate towards a dielectric surface, attach to it, and then
  propagate over the surface. The simulations are performed in air with a
  two-dimensional plasma fluid model, in which a flat dielectric is placed
  between two plate electrodes.
  Electrostatic attraction is the main mechanism that causes streamers to grow
  towards the dielectric. Due to the net charge in the streamer head, the
  dielectric gets polarized, and the electric field between the streamer and the
  dielectric is increased. Compared to streamers in bulk gas, surface streamers
  have a smaller radius, a higher electric field, a higher electron density, and
  higher propagation velocity. A higher applied voltage leads to faster
  inception and faster propagation of the surface discharge. A higher dielectric
  permittivity leads to more rapid attachment of the streamer to the surface and a thinner surface streamer. Secondary emission coefficients
  are shown to play a modest role, which is due to relatively strong
  photoionization in air. In the simulations, a high electric field is present
  between the positive streamers and the dielectric surface. We show that the
  magnitude and decay of this field are affected by the positive ion mobility.
\end{abstract}

\ioptwocol

\section{Introduction}
\label{sec:introduction}

Electric discharges in electronic devices and HV (high-voltage) equipment often
occur along dielectric materials. In the regions of HV stress around an
insulator, electron avalanches and streamer discharges can develop. These
partial discharges may eventually result in surface flashover of the insulator,
i.e., electric breakdown. In~\cite{cookson1970} it was found that around
atmospheric pressure, surface flashover voltages were 10\%-50\% lower than
flashover voltages in pure gas gaps. A dielectric present in the vicinity of the
electrodes not only modifies the fields between the electrodes, but also serves
as a possible source or sink of electrons during the breakdown process. Studying
the interaction between dielectrics and streamer discharges is therefore
important to understand surface flashover.

Early studies of surface discharges focused on the measurement of flashover voltage~\cite{cookson1981,sudarshan1986}.
In the past few decades, the use of high-speed cameras has revealed more details about the early stages of surface discharges.
In several experiments, streamer discharges were observed to have an affinity to propagate along dielectric surfaces rather than through the background gas only~\cite{sobota2008,trienekens2014}.
This affinity for a dielectric surface was reported to depend on the discharge gap geometry~\cite{sobota2009}, gas composition, pressure~\cite{dubinova2016}, and dielectric properties~\cite{allen1999,meng2015}.

To gain more insight into the physics of surface discharges, different types of numerical simulations have been performed, see e.g.~\cite{dubinova2016,jorgenson2003,jansky2010,babaeva2016,georghiou2005,sun2018,sima2016}.
Studies on the interaction between plasmas and dielectrics have often been performed at lower pressure and in noble gases, where the discharge mechanisms are relatively well understood~\cite{zhang2018b,sun2018a}. 
Several authors have also studied surface discharges in atmospheric air. An incomplete list is given below.

Jorgenson \textit{et al}.~\cite{jorgenson2003} investigated the role of photoemission in the surface breakdown process. With Monte Carlo simulations, they concluded that photoemission plays a role at low field values near the breakdown threshold.
Celestin \textit{et al}.~\cite{celestin2009} studied dielectric barrier discharges in air both experimentally and computationally, and highlighted the importance of surface charge.
J\'{a}nsk\'{y} \textit{et al}.~\cite{jansky2010} presented simulations of an air plasma discharge at atmospheric pressure, initiated by a needle anode set inside a dielectric capillary tube.
\RV{Meyer \textit{et al}.~\cite{Meyer_2019} studied surface streamers with a plasma fluid model in a 2D geometry. Agreement with empirical estimates for streamer propagation lengths was found, and it was observed that the surface charge quickly reaches so-called `saturation charge' conditions.}
Babaeva \textit{et al}.~\cite{Babaeva_2015,babaeva2016} performed a computational investigation of nanosecond pulsed surface discharges of positive and negative polarity. A hybrid fluid-Monte Carlo model was used to more accurately capture secondary electron emission caused by positive ions and photons. \xiaoran{Sima \textit{et al}.~\cite{sima2016} presented 2D axisymmetric fluid simulations of discharges spreading radially over a dielectric surface in a N$_{2}$/O$_{2}$ mixture.} Furthermore, several computational studies of plasma-liquid interaction and plasma-tissue interaction have been performed at atmospheric pressure, see for example~\cite{babaeva2013,Tian_2014}. In such studies, the liquid or skin is often modeled as a dielectric, sometimes with a finite conductivity.


The studies mentioned above have greatly improved our understanding of surface
discharges in
air. 
Here, this past work is extended in several ways. We consider a different
geometry, namely a flat dielectric placed between parallel-plate electrodes.
\RV{Streamers in this geometry can propagate along the dielectric or through the gas, in contrast to surface dielectric barrier discharges (SDBD), in which electrodes are completely separated by a dielectric \cite{babaeva2016,hua2019,brandenburg2017}. This allows us to investigate the whole streamer-dielectric interaction, including discharge inception, attachment to the dielectric and propagation over the surface.
Moreover, this geometry resembles some actual HV insulation applications \cite{meng2015, sima2017}, e.g., insulators inserted between HV and ground electrodes in gas-insulated switchgear.}
Our focus here is on positive surface streamers, which can
be computationally expensive to simulate. We have therefore developed an
efficient fluid model with adaptive mesh refinement. For simplicity and
efficiency, 2D simulations are used here, as full 3D simulations would still be
very costly. 

The content of the paper is as follows. The simulation model is described in
section \ref{sec:simulation-model}. In section \ref{sec:interaction-dielectric},
we focus on the attraction of streamers to dielectrics, and we look at the
differences between surface and gas-phase streamers. Afterwards, several discharge parameters are varied, to study their effect on the streamer's inception time, propagation
velocity and morphology:
\begin{itemize}
  \item The applied voltage in section \ref{sec:effect-appl-volt}
  \item The permittivity ($\varepsilon$) of the dielectric material in section
  \ref{sec:effect-perm-diel}
  \item The secondary electron emission coefficients (for positive ions and photons) in section \ref{sec:effect-electron-emission}
  \item The positive ion mobility in section~\ref{sec:effect-positive-ion}.
\end{itemize}

\section{Simulation Model}
\label{sec:simulation-model}

The 2D fluid model used in this paper is based on
Afivo-streamer~\cite{teunissen2017,teunissen2018}, which is an open-source
plasma fluid code for streamer discharges that features adaptive mesh refinement
(AMR), geometric multigrid methods for Poisson's equation, and OpenMP
parallelism. For a recent comparison of six streamer simulation codes, including
Afivo-streamer, see~\cite{bagheri2018}. We have made several modifications to be
able to simulate surface streamers:
\begin{itemize}
  \item Electrons, ions and photons can be absorbed by dielectric surfaces.
  \item Surface densities and fluxes are stored separately from their equivalents in the gas.
  \item The electric field computation takes the surface charge into account.
  \item A new Monte Carlo photoemission module was implemented, and the
  photoionization routines were adjusted to account for the dielectric.
\end{itemize}
These changes are described in more detail below.

With our 2D model, we effective simulate planar surface discharges. This leads
to some differences compared to a full 3D description. First, the electric
fields and charge densities in 3D are typically higher, as the streamer heads
have a stronger curvature. Second, it is often observed that both surface and
gas streamers are present in experiments~\cite{allen1999,meng2017}. We do not
observe these two components in our 2D model, but have seen them in preliminary
3D simulations that are still under development.

\subsection{Fluid Model}
\label{sec:fluid-model}

The fluid model used here is of the drift-diffusion-reaction type with the local field approximation~\cite{luque2012}.
The model keeps track of the electron density \textit{n}$_e$, the positive ion density \textit{n}$_{i}^{+}$ and the negative ion density \textit{n}$_{i}^{-}$, which involve in time as
\begin{eqnarray}
  \frac{\partial n_e}{\partial t}= -\nabla \cdot \vec{\Gamma}_e + S_{i}-S_a + S_{pi}+\RV{S_{se}},\label{eq:electron-ddt}\\
  \qquad \vec{\Gamma}_e = -n_e\mu_e \vec{E} - D_e\nabla n_e,\nonumber\\
  \frac{\partial {n_{i}}^{+}}{\partial t}=-\nabla \cdot \vec{\Gamma}_i^+ + S_{i}+S_{pi},\\
  \qquad \vec{\Gamma}_i^+ = {n_{i}}^{+}\mu_i^{+} \vec{E},\nonumber\\
  \frac{\partial {n_{i}}^{-}}{\partial t}=-\nabla \cdot \vec{\Gamma}_i^-+S_{a},\\
  \qquad \vec{\Gamma}_i^- = -{n_{i}}^{-} \mu_i^{-} \vec{E},\nonumber
\end{eqnarray}
Here, fluxes are indicated by a $\vec{\Gamma}$, ${\mu}_e$ is the electron mobility, $D_e$ the electron diffusion
coefficient, $\vec{E}$ the electric field, and $\mu_i^\pm$ the positive/negative ion mobilities.
Furthermore, several source terms are present. The electron impact ionization
and electron attachment terms are given by $S_i = \alpha \mu_e |\vec{E}| n_e$
and $S_a = \eta \mu_e |\vec{E}| n_e$, respectively, where $\alpha$ and $\eta$
are the ionization and attachment coefficients. The production of photoelectrons
from photoionization is included with the term $S_{pi}$. \RV{Secondary electron
emission due to the impact of both ions and photons is accounted for by the term $S_{se}$, defined below.}

The local field approximation is used, so that ${\mu}_e$, $D_e$, $\alpha$ and
$\eta$ are functions of the local electric field strength. Electron transport
and reaction coefficients for air ($1 \, \textrm{bar}$, $300 \, \textrm{K}$)
were generated with Monte Carlo particle swarm simulations (see
e.g.~\cite{Rabie_2016a}), using Phelps' cross sections~\cite{Phelps_1985}. The positive
ion mobility $\mu_i^+ = 3 \times 10^{-4}\,\mathrm{m}^{2}/\mathrm{Vs}$ is here
considered to be constant, but in section \ref{sec:effect-positive-ion} it is varied to investigate its effect on surface discharges. For simplicity, the negative ion mobility is set to zero ($\mu_i^- = 0$) throughout the paper.

We assume that electrons and ions attach to the surface when they flow onto a
dielectric. They do not
move or react on the surface, but secondary electron emission from the surface
is taken into account. For the impact of positive ions, a SEE (secondary
electron emission) coefficient ${\gamma}_{i}$ is used. When a photon hits a
dielectric surface, we assume that the photon is absorbed. A SEE
photoemission coefficient ${\gamma}_{pe}$ is used \RV{to determine the photoemission flux, see section \ref{sec:photoionization-emission}.} The effect of these SEE coefficients is studied in section \ref{sec:effect-electron-emission}, elsewhere they are set to zero.
\RV{The secondary electron emission source term $S_{se}$ in equation (\ref{eq:electron-ddt}) is non-zero only in cells adjacent to the dielectric surface. In these cells, it is given by
  \begin{equation}
    S_{se} = -\nabla \cdot (\tilde\vec{\Gamma}_{pe} - \gamma_{i} \tilde\vec{\Gamma}_{i}^{+}),
  \end{equation}
where $\tilde{\Gamma}_{i}^{+}$ is flux of positive ions onto the surface, and $\tilde{\Gamma}_{pe}$ is the photoemission flux coming from the surface. By definition, both these fluxes are non-zero only at the surface.} 

Secondary emission leaves
behind positive surface charge on the dielectric. Therefore, the surface charge
density ${\sigma}_{s}$ changes in time as
\begin{equation}
  \partial_t \sigma_{s}=-e(\tilde{\Gamma}_e+\xiaoran{\tilde{\Gamma}_i^{-}})
  + e(1+\gamma_{i})\tilde{\Gamma}_{i}^{+} + \RV{e\tilde{\Gamma}_{pe}},
\end{equation}
where $e$ is the elementary charge and the other terms correspond to the fluxes
onto the dielectric surface: $\tilde{\Gamma}_e$ for electrons, \xiaoran{$\tilde{\Gamma}_{i}^{-}$ for
	negative ions,}  $\tilde{\Gamma}_{i}^{+}$ for positive ions, and $\tilde{\Gamma}_{pe}$ for
photons. We study positive streamers, which means that electrons generally move away from dielectrics. Therefore, an accurate description of the electron flux towards the surface~\cite{Hagelaar_2000} is not required here.

\subsection{Electric Field}
\label{sec:electric-field}

The electric field $\vec{E}$ is calculated by first solving Poisson's equation for the electric potential $\varphi$:
\begin{eqnarray}
\nabla \cdot \left(\varepsilon \nabla \varphi \right)=-(\rho +\delta_{s}\sigma_{s}),
\end{eqnarray}
where $\varepsilon$ is the dielectric permittivity, ${\rho}$ is the volume
charge density, and $\delta_{s}$ maps the surface charge $\sigma_s$ on the
gas-dielectric interface to the grid cells adjacent to the dielectric. Afterwards,
the electric field is computed as
\begin{equation}
\vec{E}=-\nabla \varphi.
\label{equ:solve E}
\end{equation}
At the dielectric interface, we ensure that the normal component of the
electric field satisfies the classic jump condition
\begin{equation}
  \varepsilon_{2}E_{2}- \varepsilon_{1}E_{1}=\sigma_{s},
\label{equ:solve E at boundary}
\end{equation}
\RV{where $\varepsilon_1$ and $\varepsilon_1$ denote the permittivities on both sides of the interface, and $E_1$ and $E_2$ the electric field components normal to the interface.}

Details about the numerical implementation, which is compatible with adaptive
mesh refinement, will be presented in a forthcoming paper.

\subsection{Photoionization and Photoemission}
\label{sec:photoionization-emission}

\begin{figure}
  \centering
  \includegraphics[width=1.0\linewidth]{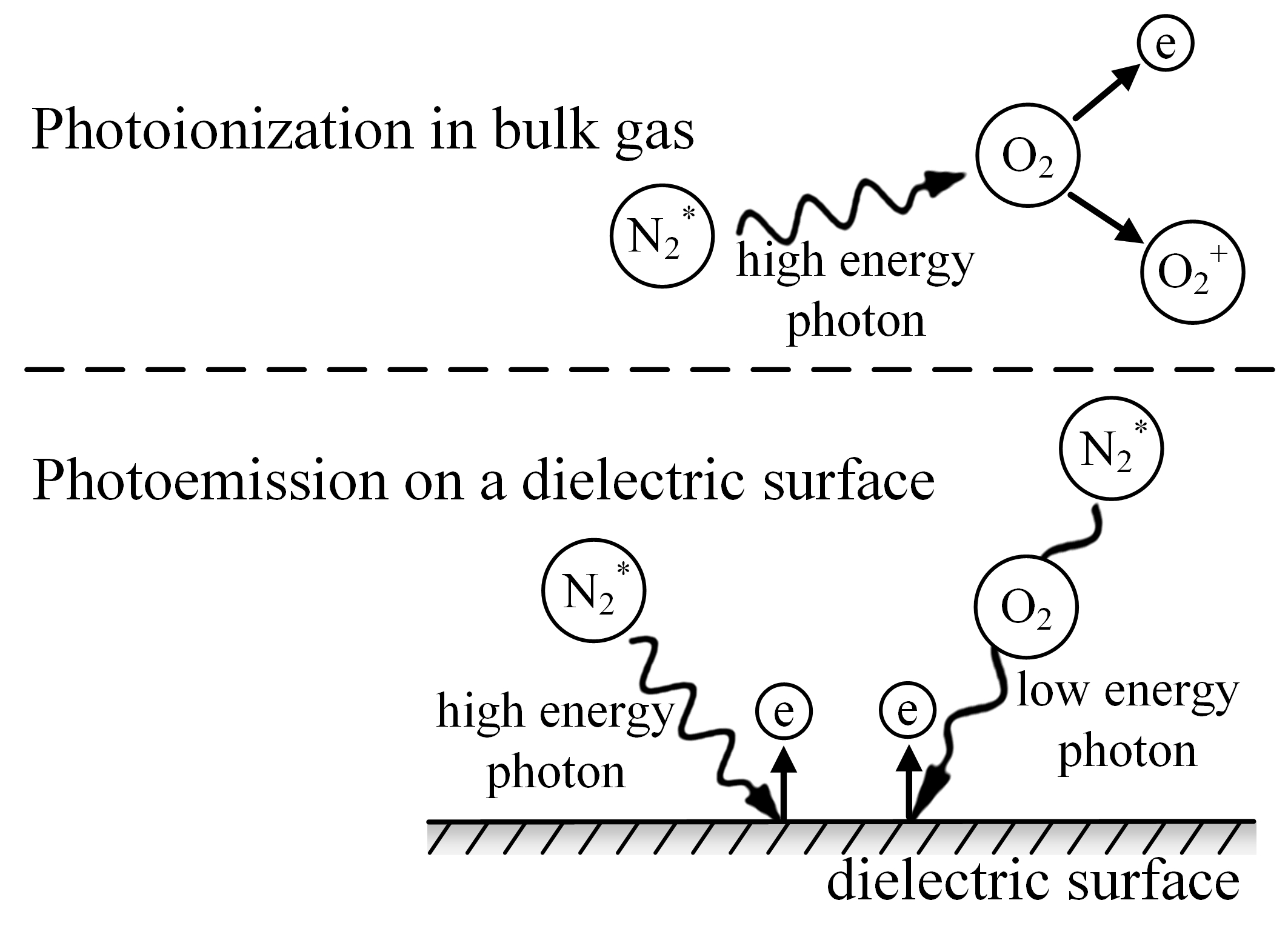}
  \caption{Illustration of photoionization and photoemission mechanisms in air.
    Two types of photons are considered. High-energy photons can generate
    photoionization and photoemission, whereas low-energy photons are not
    absorbed by the gas and only contribute to photoemission. }
  \label{fig:photo-processes}
\end{figure}

Positive streamer discharges need a source of free electrons ahead of them in order to propagate.
Photons can generate such free electrons through photoionization in the gas or photoemission from a dielectric surface.
In $\mathrm{N}_{2}$--$\mathrm{O}_{2}$ mixtures, non-local photoionization can take place when an excited nitrogen molecule emits a UV photon in the 98 to 102.5 nm range, which has enough energy to ionize an oxygen molecule. Photoionization often plays an important role in electrical discharges, see e.g.~\cite{pancheshnyi2005,Pancheshnyi_2014}.

The role of photoemission in surface discharges is less well understood. Photons
can be emitted from several excited states. The probability of photoemission not
only depends on the photon energy, but also on the surface
properties~\cite{jorgenson2003}. For simplicity, we consider only two types of
photons in this paper: high-energy photons, which can generate photoionization
and photoemission, and low-energy photons, which can only contribute to
photoemission and are not absorbed in the gas. These processes are illustrated
in figure \ref{fig:photo-processes}.

\RV{Photoionization and photoemission are here modeled with a Monte Carlo (MC)
  method. We use the same Monte Carlo photoionization model as described in \cite{Bagheri_2019} and chapter 11 of \cite{teunissen}. The idea is to approximate the photoionization source term $S_{pi}$ and the photon flux onto the dielectric $\tilde{\Gamma}_{pe}$ by randomly sampling discrete photons. For computational efficiency, these terms are updated every $\Delta t_\gamma = 10 \Delta t$, where $\Delta t$ is the time step used for solving equation (\ref{eq:electron-ddt}).}

\RV{First, the number of ionizing (high-energy) photons produced per grid cell during a time $\Delta t_\gamma$ is determined. We use Zheleznyak's model~\cite{zheleznyak1982}, in which the number of ionizing photons is proportional to the number of impact ionization events.
The corresponding proportionality factor is set to $\xi p_q / (p + p_q)$, where
$\xi = 0.05$ is a numerical factor, $p = 1 \, \textrm{bar}$ is the gas
pressure and $p_q = 40 \, \textrm{mbar}$ is the collisional quenching pressure. We remark that $\xi$ should in principle depend on the electric field~\cite{zheleznyak1982}, but that it is here approximated by a constant, as in~\cite{Bagheri_2019}. Per cell, a random number is drawn to determine how many photons are generated, see \cite{teunissen}.}

\RV{Since simulations are here performed in 2D, the discrete photons do not
correspond to (single) physical photons. Instead, the total photon number $N_\mathrm{photons}$ is
fixed, so that the MC method always uses $10^5$ photons. The weight factor $w$ of these photons is given by $w = \int S_\gamma dV / N_\mathrm{photons}$, where
\begin{equation}
  \int S_\gamma dV = \xi p_q / (p + p_q) \int S_i dV,
\end{equation}
is the volume-integrated production rate of ionizing photons.}

\RV{For simplicity, the number of produced low-energy photons is assumed to be equal to the number of high-energy
photons. As the low-energy photons only contribute to photoemission, their
effect can be controlled through the corresponding photoemission coefficient.}

\RV{Second, an isotropically distributed direction is sampled for each photon. Afterwards, absorption lengths are determined. The absorption length of high-energy photons is sampled from the absorption function
for air, see e.g.~\cite{Bagheri_2019}. Low-energy photons are not absorbed by the gas, so their absorption length is set to a large value, making sure they always end up outside the computational domain.}

\RV{Third, we determine which photons hit a dielectric surface, and where they do so. These photons are absorbed by the surface, where they contribute to the local photoemission flux $\tilde{\Gamma}_{pe}$. For the
low-energy and high-energy photons, photoemission coefficients ${\gamma}_{peL}$
and ${\gamma}_{peH}$ are used, respectively. If a surface cell of area $\Delta A$ is hit by $n_{L}$ and $n_{H}$ low-energy and high-energy photons, then
\begin{equation}
  \tilde{\Gamma}_{pe} = ({\gamma}_{peL} n_{L} + {\gamma}_{peH} n_{H}) w / \Delta A.
\end{equation}
The effect of photoemission is
investigated in section \ref{sec:effect-electron-emission}; elsewhere in the
paper photoemission is not taken into account (so that
${\gamma}_{peL} = {\gamma}_{peH} = 0$).}

\RV{Fourth, the remaining high-energy photons that are absorbed in the gas
  contribute to photoionization source term $S_{pi}$. If $n_\gamma$ photons are
  absorbed in a grid cell with volume $\Delta V$, then
  \begin{equation}
  S_{pi} = n_\gamma w / \Delta V.
  \end{equation}
  Low-energy photons and high-energy photons
  that are absorbed outside the computational domain have no effect.}

\subsection{Computational domain and initial conditions}
\label{sec:parameters}

\begin{figure}
  \centering
  \includegraphics[width=0.6\linewidth]{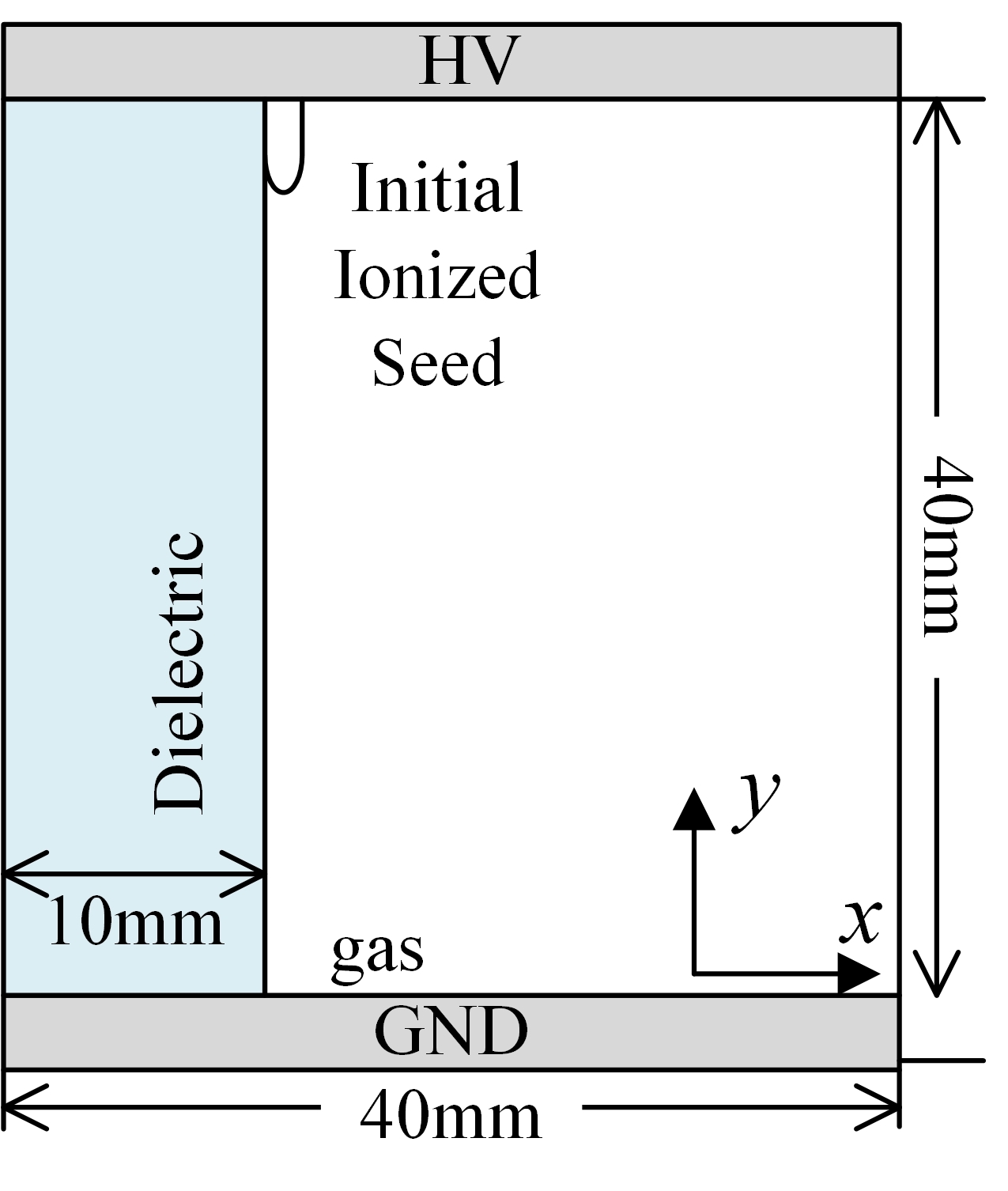}
  \caption{The computational domain. A parallel-plate geometry is used, with a
    flat dielectric present on the left. Discharges start from the ionized seed
    present close to the top electrode, as described in the text.}
  \label{fig:figure-2}
\end{figure}

We use a parallel-plate electrode geometry with a flat dielectric in between, as
shown in Figure \ref{fig:figure-2}. The computational domain measures (40
mm)$^{2}$, and a dielectric is present on the left side with a width of $10 \, \mathrm{mm}$. The
dielectric permittivity is set to $\varepsilon = 2$, but in section
\ref{sec:effect-perm-diel} it is varied to investigate
its effect on surface discharges.

As a gas, artificial air (80\% N$_{2}$ and 20\% O$_{2}$) at 1 bar and 300 K is
used. For the electric potential, Dirichlet boundary conditions are applied at
the upper and lower boundaries, and Neumann zero boundary conditions on the left
and right side. A voltage of $100 \, \textrm{kV}$ is applied. In section
\ref{sec:effect-appl-volt}, this voltage is varied. The background densities of
electrons and positive ions are set to 10$^{10}$ m$^{{-}3}$~\cite{sun2014}.

To start a discharge, the background field has to be locally enhanced. We do
this by placing an ionized seed of about $2 \, \mathrm{mm}$ long with a radius of about 0.4
mm. The electron and positive ion density are 5${\times}$10$^{18}$ m$^{{-}3}$ at
the center, and they decay at distances above $d=0.2 \, \mathrm{mm}$ with a
so-called smoothstep profile: $1-3x^{2}+2x^{3}$ up to $x = 1$, where
$x = (d-0.2 \, \mathrm{mm})/0.2 \, \mathrm{mm}$. When the electrons from a seed
drift upwards, the electric field at the bottom of the seed is enhanced so that
a positive streamer can form.

\section{Results \& discussion}
\label{sec:results-discussion}


\begin{table*}
	\centering
	\begin{tabular}{ccccccc}
          \hline
          Section & d (mm) & $U$ (kV) & $\varepsilon_r$ & $\gamma_{i}$ &$\gamma_{pe}$ & $\mu_i^+$ (m$^{2}$/Vs)  \\ \hline
          3.1 & $(0.5,1,2,5)$ & 100 & 2 & 0 & 0 & $3\times10^{-4}$\\
          3.2 & 0.5 & $(92,100,112)$ & 2 & 0 & 0 & $3\times10^{-4}$ \\
          3.3 & 0.5 & 100 & $(2,3,5)$ & 0 & 0 & $3\times10^{-4}$ \\
          3.4.1 & 0.4 & 100 & 2 & $(0,0.5)$ & 0 & $3\times10^{-4}$ \\
          3.4.2 & $(0.5,1)$ & 100 & 2 & 0 & $(0,0.5)$ & $3\times10^{-4}$ \\
          3.5 & 0.5 & 100 & 2 & 0 & 0 & $(0, 1, 5, 10) \times10^{-4}$\\
          \hline
	\end{tabular}
        \caption{Investigated parameters and their values in each section. Here
          $d$ is the distance between seed center and the dielectric; $U$ the
          applied voltage; $\epsilon_{r}$ the relative permittivity of the
          dielectric; $\gamma_{i}$ the ion-induced secondary electron emission
          coefficient; $\gamma_{pe}$ the photoemission coefficient and
          $\mu_i^+$ the positive ion mobility.} 
        \label{parameter_table}
\end{table*}

In section \ref{sec:interaction-dielectric}, the initial seed is placed at
different distances from the dielectric to study the streamer-dielectric
interaction. We also point out the main differences between \emph{surface} and
\emph{gas} streamers. Next, we systematically study the effect of several
parameters on the surface discharges: the applied voltage (section
\ref{sec:effect-appl-volt}), the dielectric permittivity (section
\ref{sec:effect-perm-diel}), the secondary electron emission coefficients
(section \ref{sec:effect-electron-emission}), and finally the ion mobility
(section \ref{sec:effect-positive-ion}).
\xiaoran{The parameters investigated and their values in each section are shown in table \ref{parameter_table}.}

\subsection{Streamer-dielectric interaction}
\label{sec:interaction-dielectric}

\begin{figure}
	\centering
	\includegraphics[width=1.0\linewidth]{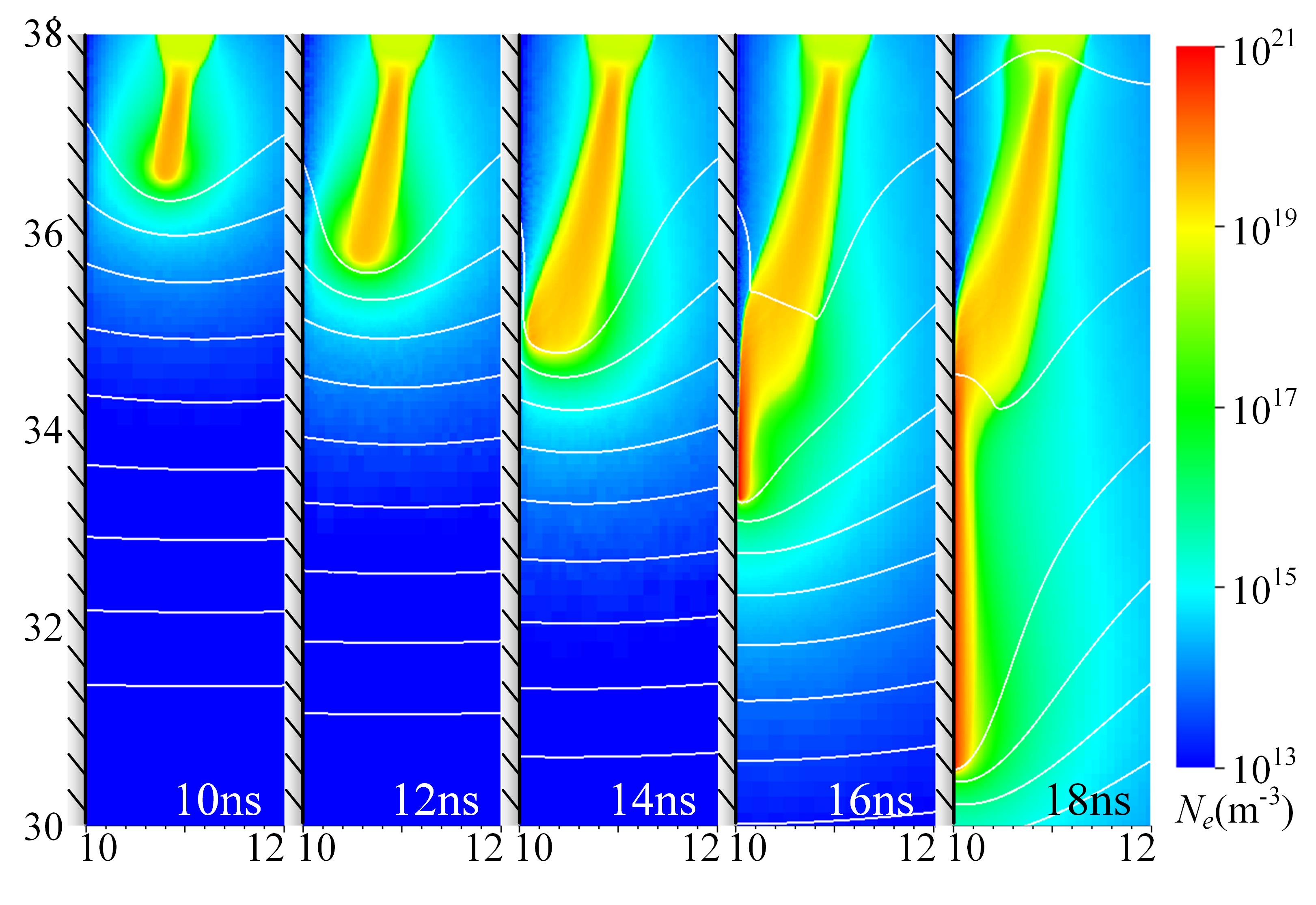}
	\caption{The streamer development process between $10 \, \mathrm{ns}$ and $18 \, \mathrm{ns}$ for seed placed at $1 \, \mathrm{mm}$ from the dielectric. \RV{White equipotential lines spaced by 2 kV are shown in part of the domain.}}
	\label{fig:process_d1}
\end{figure}

Previous experiments have revealed that dielectrics attract positive
streamers, see e.g.~\cite{sobota2008,trienekens2014}.
\xiaoran{This attraction is also present in our numerical model. Figure \ref{fig:process_d1} shows the evolution of the electron density for an initial ionized seed placed $1 \, \mathrm{mm}$ away from the dielectric.
	It can be seen that the streamer start to grow in air and then gradually develops towards the dielectric. After connecting with the dielectric, the streamer propagates down over its surface. The evolution shown here is in qualitative agreement with the discharge photographs in \cite{sobota2008}.}

\begin{figure}
  \centering
  \includegraphics[width=1.0\linewidth]{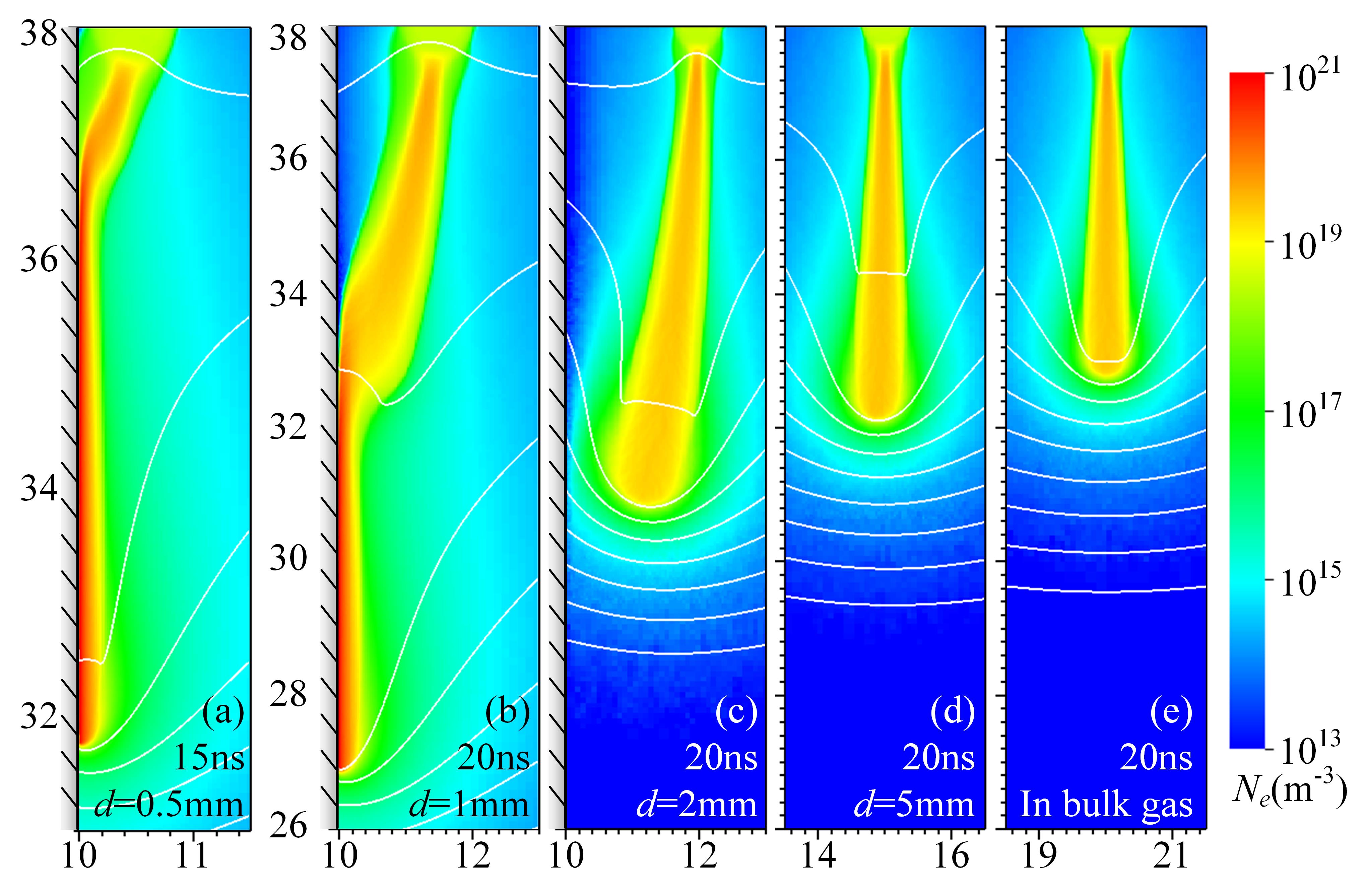}
  \caption{The electron density for streamers starting from
    different locations. For panels a--d, the initial seed was placed at $0.5 \, \mathrm{mm}$,
    $1 \, \mathrm{mm}$, $2 \, \mathrm{mm}$ and $5 \, \mathrm{mm}$ from the dielectric. For comparison, a streamer in bulk gas
    is shown in panel e. Results are shown at $20 \, \mathrm{ns}$, except for panel a, which
    has the fastest propagation. \RV{White equipotential lines spaced by 2 kV are shown in part of the domain.}}
  \label{fig:diff seed position}
\end{figure}

To study the streamer-dielectric attraction, we have placed the initial ionized seed at different distances from the dielectric.
\xiaoran{Figures \ref{fig:diff seed position}a--d show the electron density for seeds placed at $0.5 \, \mathrm{mm}$, $1 \, \mathrm{mm}$, $2 \, \mathrm{mm}$ and $5 \, \mathrm{mm}$ from the dielectric.
	\RV{For comparison, the electron density for a streamer in
	bulk gas is also shown in figure \ref{fig:diff seed position}e.
In this simulation the dielectric was removed, so that the whole computational domain contained gas, and the initial seed was placed at $x = 20 \, \textrm{mm}$.}
	 Figure \ref{fig:diff seed position}a shows the electron density at $15 \, \mathrm{ns}$, for the other cases, which develop more slowly, results at $20 \, \mathrm{ns}$ are shown.}
The closer the streamer is located to the dielectric,
the stronger the attraction to the dielectric becomes. It can also be seen that a nearby dielectric increases the streamer's velocity in the gas, and that streamers here propagate faster on the surface than in the gas.

\subsubsection{Attraction to the dielectric}

\begin{figure*}
  \centering
  \includegraphics[width=0.9\linewidth]{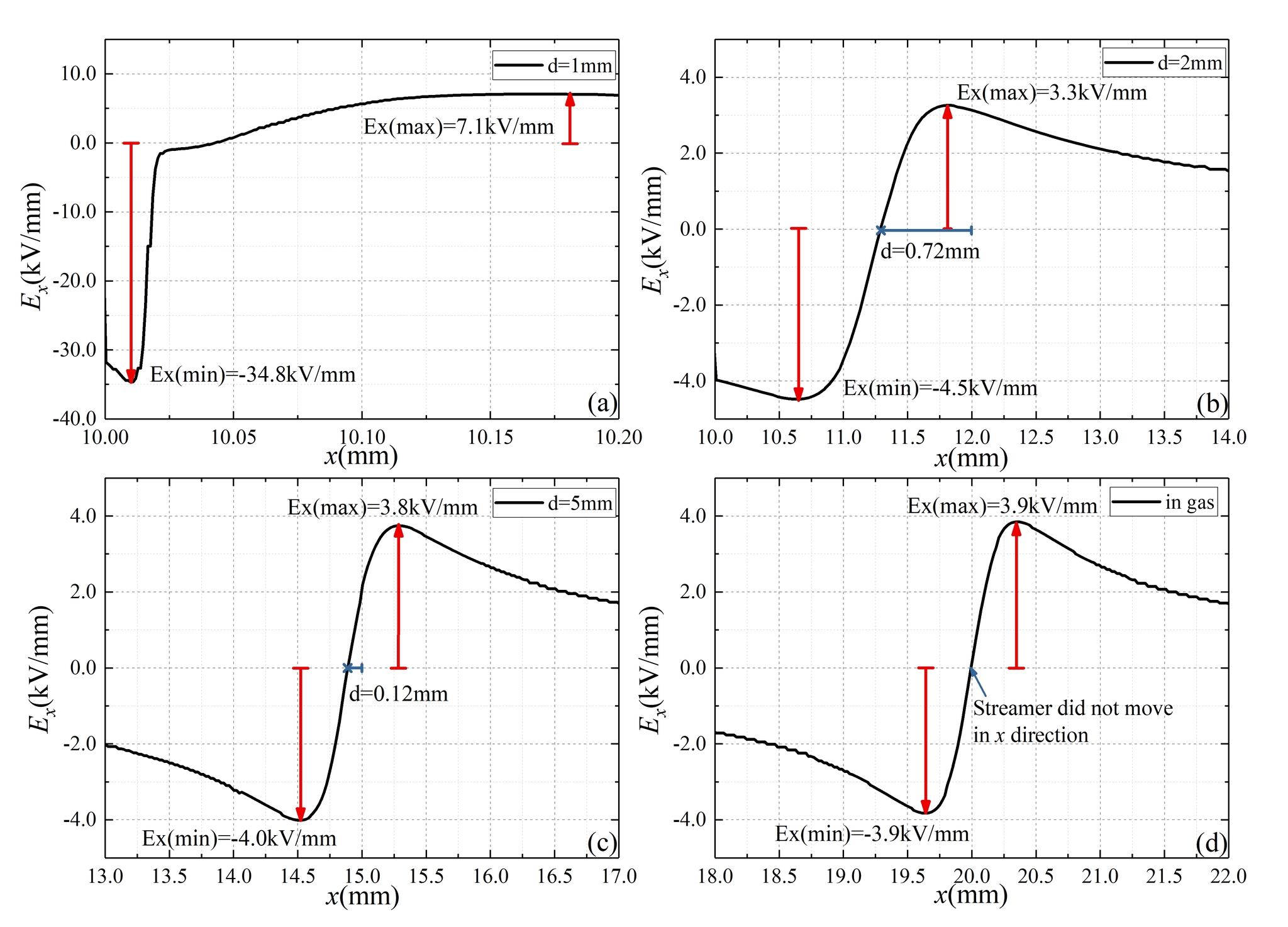}
  \caption{Horizontal electric field ($E_{x}$) at $20 \, \mathrm{ns}$ for streamers starting from different locations. The curves are taken at the streamer head, at the height of the maximum electric field.}
  \label{fig:Ex for diff seeds}
\end{figure*}

As photoemission is disabled here (see table \ref{parameter_table}), the attraction of the streamer to the dielectric is purely electrostatic. The net charge in the streamer head polarizes the dielectric, which increases the electric field between the streamer and the dielectric.
This effect is illustrated in figure \ref{fig:Ex for diff seeds}, which shows the horizontal electric field ($E_{x}$) around the streamer heads. With a dielectric present, $|E_{x}|$ increases on the dielectric side, and it is reduced on the other side.
The closer the streamer is to a dielectric, stronger this effect becomes. Eventually, the streamer will turn into a surface streamer. As shown in figure \ref{fig:diff seed position}, such a surface streamer is thinner and quite asymmetric compared to a gas streamer.

\RV{In past research, the attraction of streamers to dielectrics could often be
  explained by the enhanced static field between pointed electrodes and
  dielectrics, see e.g.~\cite{sobota2008,sobota2009}. In contrast, the
  attraction to the dielectric is here due to the space charge from the streamer
  itself, since in our plate-to-plate geometry, the electric field has no
  horizontal component before a discharge is present. Assuming that streamers
  propagate approximately along electric field lines~\cite{Nijdam_2014}, we can
  therefore state that streamers lead themselves to the dielectric: their space
  charge modifies the background field and causes a horizontal field component
  that attracts them to the dielectric.}

\RV{Surface charge on the dielectric could also play a role in attracting
  streamer discharges, by modifying the background electric field. However,
  there is negligible surface charge here, as secondary electron emission is
  disabled in this section and electrons move away from the dielectric.
  Positive ions do move towards the dielectric, but there are initially
  few of them near the dielectric, and they drift with a relatively low mobility, so
  that they hardly contribute to the surface charge.}

\subsubsection{Effect on streamer velocity}

\begin{figure}
  \centering
  \includegraphics[width=0.9\linewidth]{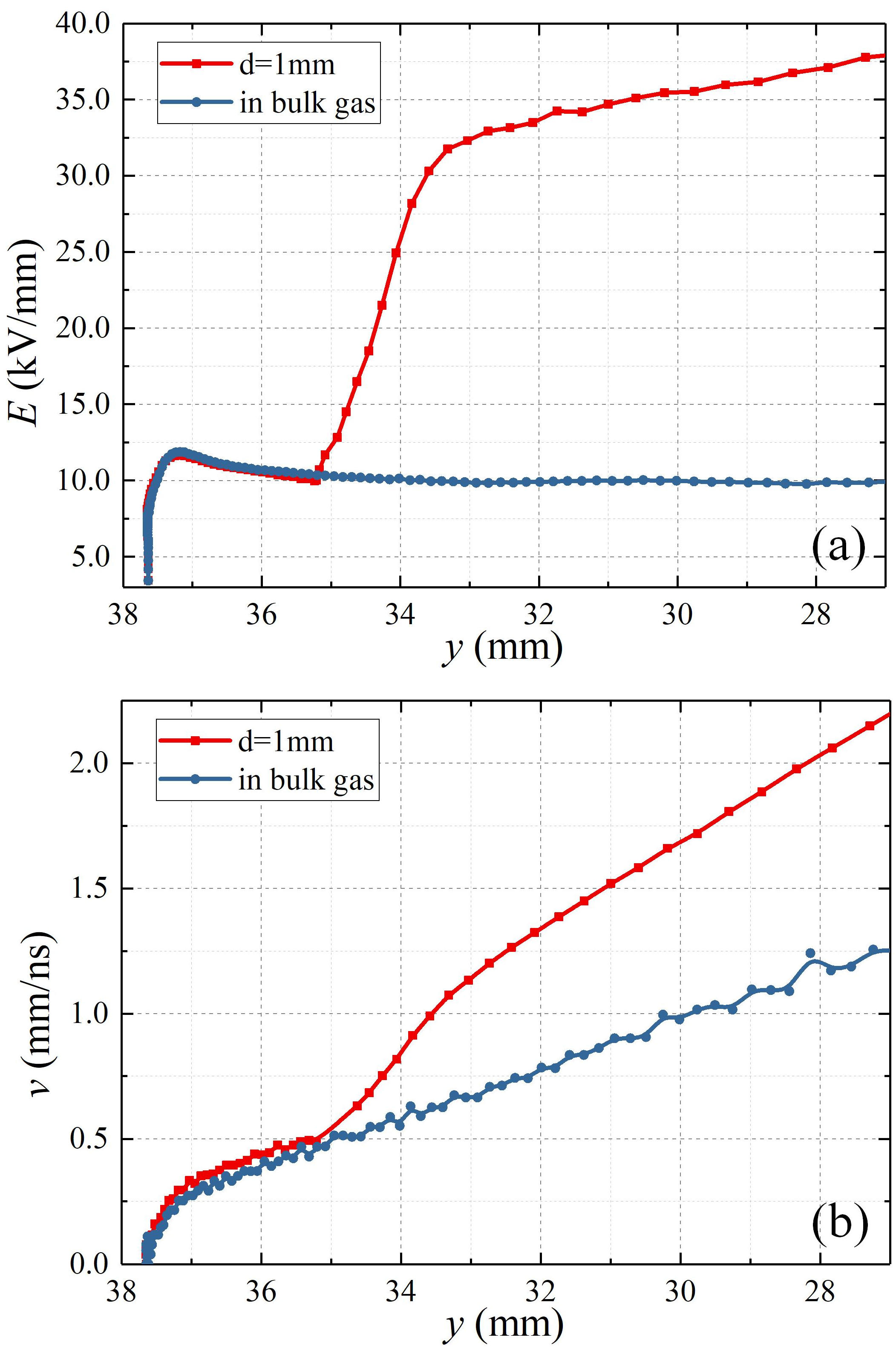}
  \caption{The maximum electric field and streamer velocity versus the vertical
    position of the streamer head.}
  \label{fig:v-maxE-y}
\end{figure}

Several experimental studies have found that surface discharges are faster than bulk gas streamers~\cite{trienekens2014,allen1999,akyuz2001}. Their increased velocity was attributed to increased ionization rates near the dielectric, accumulated negative charge and electron emission from the dielectric surface.
Electron emission from the dielectric is not included here (it is in section \ref{sec:effect-electron-emission}), but we still find that the surface streamers are significantly faster. Figure \ref{fig:v-maxE-y} shows the streamer velocity and its maximal electric field for the case $d = 1 \, \textrm{mm}$. Note that both the velocity and the maximal electric field increase when the surface streamer forms, at around $y = 35 \, \textrm{mm}$. Even though the higher field is mostly in the horizontal direction, see figure \ref{fig:Ex for diff seeds}, it still contributes to a faster vertical growth.

As can be seen in figure \ref{fig:diff seed position}, the electron density
inside a surface streamer \RV{($\sim 10^{20}$ to $10^{21} \, \mathrm{m}^{-3}$)} is higher than in
a gas streamer \RV{($\sim 10^{19} \, \mathrm{m}^{-3}$) when the discharge conditions are otherwise similar. This was also observed in for example~\cite{hua2019}. We remark that besides the presence of a dielectric, the electron density inside a streamer discharge also depends on other factors, such as the applied voltage, the streamer radius, the gas composition, the amount of photoionization etc. Also note that we use a Cartesian 2D model, in which there is less field enhancement around the streamer heads than in full 3D, reducing the electron density in both surface and gas streamers.}

There seem to be several
related effects that lead to the increased surface streamer velocity. The strong
electric field between a surface streamers and a dielectric pulls surface
streamers towards the dielectric. This reduces their radius \xiaoran{(as shown in figure \ref{fig:diff seed position})}, and results in an
asymmetric streamer head shape, which also leads to stronger electric field
enhancement. The result is that the ionization rate is increased, that the
streamer has a higher degree of ionization, and that it propagates faster. This
behavior is quite distinct from gas streamers, which typically propagate faster
when they have a larger radius~\cite{Briels_2008}.

\subsubsection{Cathode sheath}
\label{sec:cathode-sheath}

\begin{figure}
	\centering
	\includegraphics[width=0.95\linewidth]{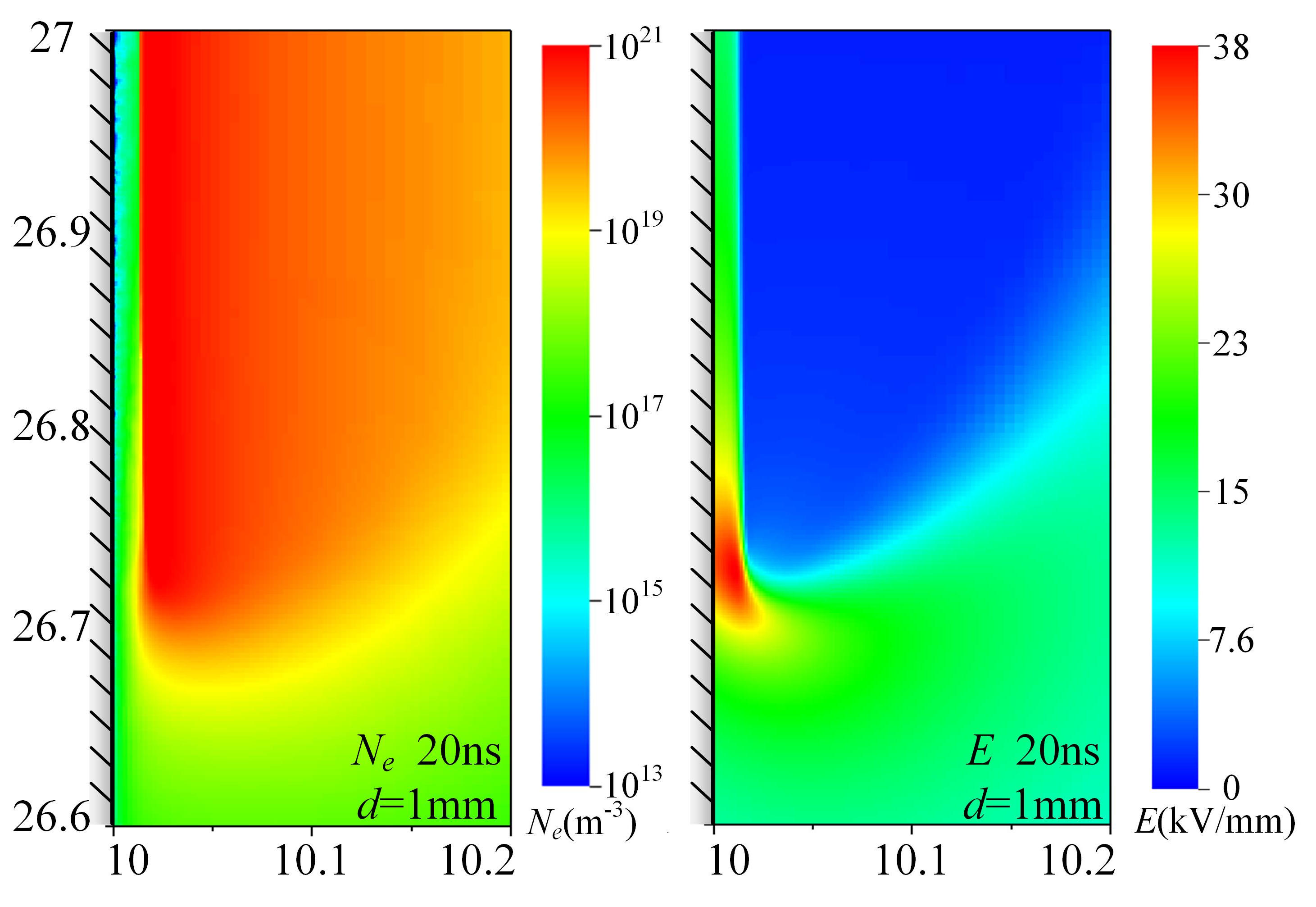}
	\caption{The electron density and the electric field around the positive streamer head at $20 \, \mathrm{ns}$, for an initial seed at $1 \, \mathrm{mm}$ from the dielectric. Note the gap between the streamer and the dielectric.}
	\label{fig:floating_streamer}
\end{figure}

As shown in figure \ref{fig:floating_streamer}, the surface streamer `hovers'
over the dielectric surface without fully connecting to it. This phenomenon was
also observed in simulations of dielectric barrier
discharges~\cite{babaeva2011,babaeva2016,stepanyan2014,soloviev2017}, and it only occurs for positive streamers. The reason is that positive streamers grow from incoming electron avalanches, but such avalanches require sufficient distance before they reach ionization levels comparable to the discharge. Positive streamers can therefore not immediately connect to the dielectric surface. Due to the net charge in the streamer head, a very high electric field is present in the narrow gap between streamer and dielectric.
The effect of secondary electron emission on these dynamics is studied in section \ref{sec:effect-electron-emission}, and the role of the positive ion
mobility is investigated in section \ref{sec:effect-positive-ion}.

We remark that the maximum electric field of a positive surface streamer can
rapidly rise to very high values, as shown in figures \ref{fig:v-maxE-y}a and \ref{fig:floating_streamer}. These
high-field areas typically contain a low electron density, but they still
pose a problem for plasma fluid simulations. \RV{In the present fluid simulations, the transport and reaction coefficients (e.g., the ionization coefficient or the electron mobility) are functions of the local electric field strength. These coefficients are tabulated up to a certain maximum electric field, which is here $35 \, \mathrm{kV/mm}$; for higher fields, we use the tabulated value at $35 \, \mathrm{kV/mm}$.} More generally, the validity of the local field approximation is
questionable when there are such high electric fields (and corresponding strong
gradients). For future studies in such ultra-high
electric fields, particle-in-cell simulations could therefore be more suitable, as was also observed in~\cite{Babaeva_2015}.
Finally, we remark that a potential physical limitation for this maximum
electric field is field emission of electrons from the surface.

\subsection{Effect of applied voltage}
\label{sec:effect-appl-volt}

\begin{figure}
	\centering
	\includegraphics[width=1.0\linewidth]{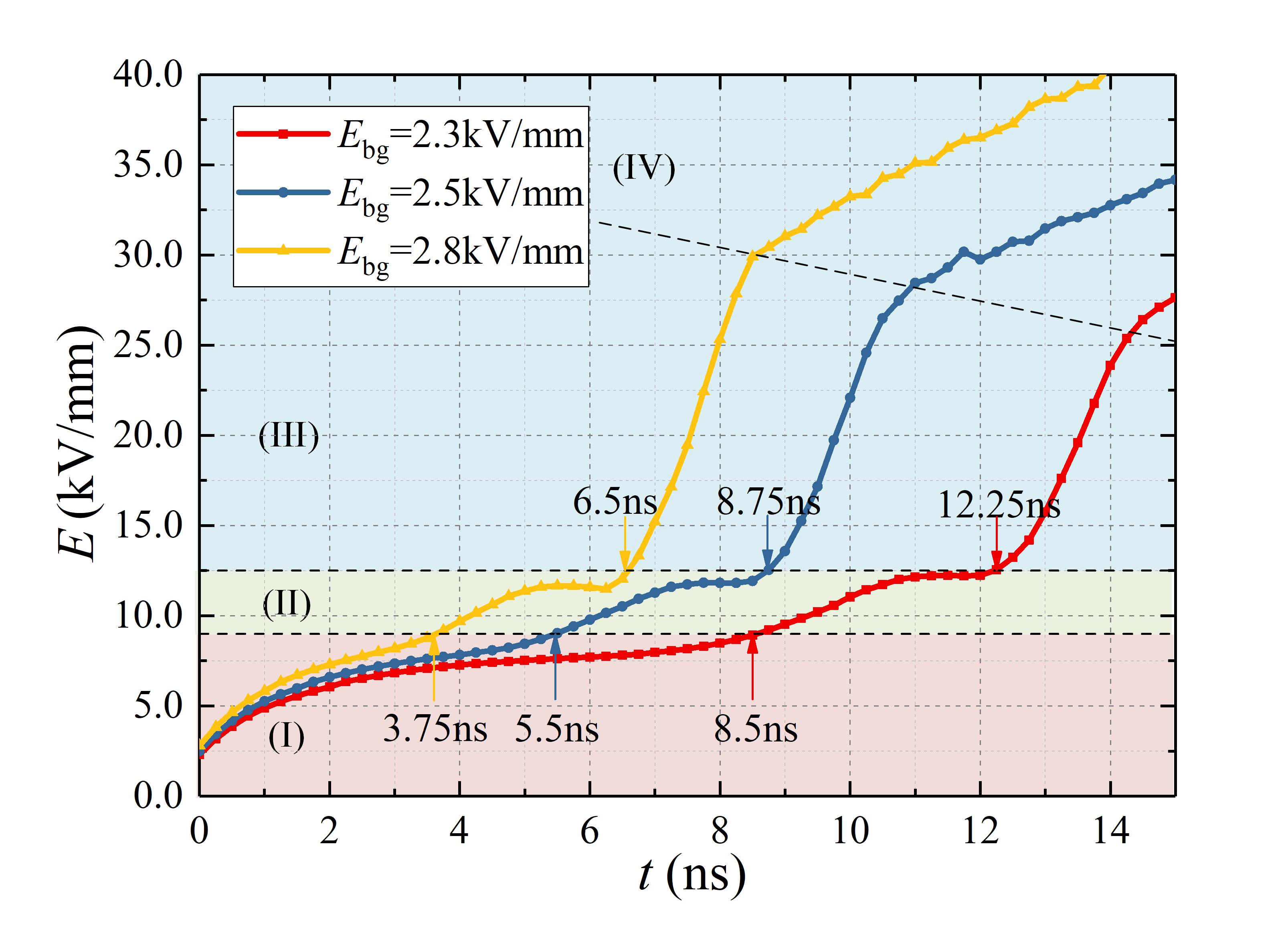}
	\caption{The maximum electric field versus time for streamers in background electric fields of $2.3 \, \mathrm{kV/mm}$, $2.5 \, \mathrm{kV/mm}$ and $2.8 \, \mathrm{kV/mm}$ between. The indicated stages are I: initial stage, II: gas propagation, III: transition towards a surface streamer, IV: stable surface propagation.}
	\label{fig:maxE_t_diffE}
\end{figure}

\begin{figure}
	\centering
	\includegraphics[width=1.0\linewidth]{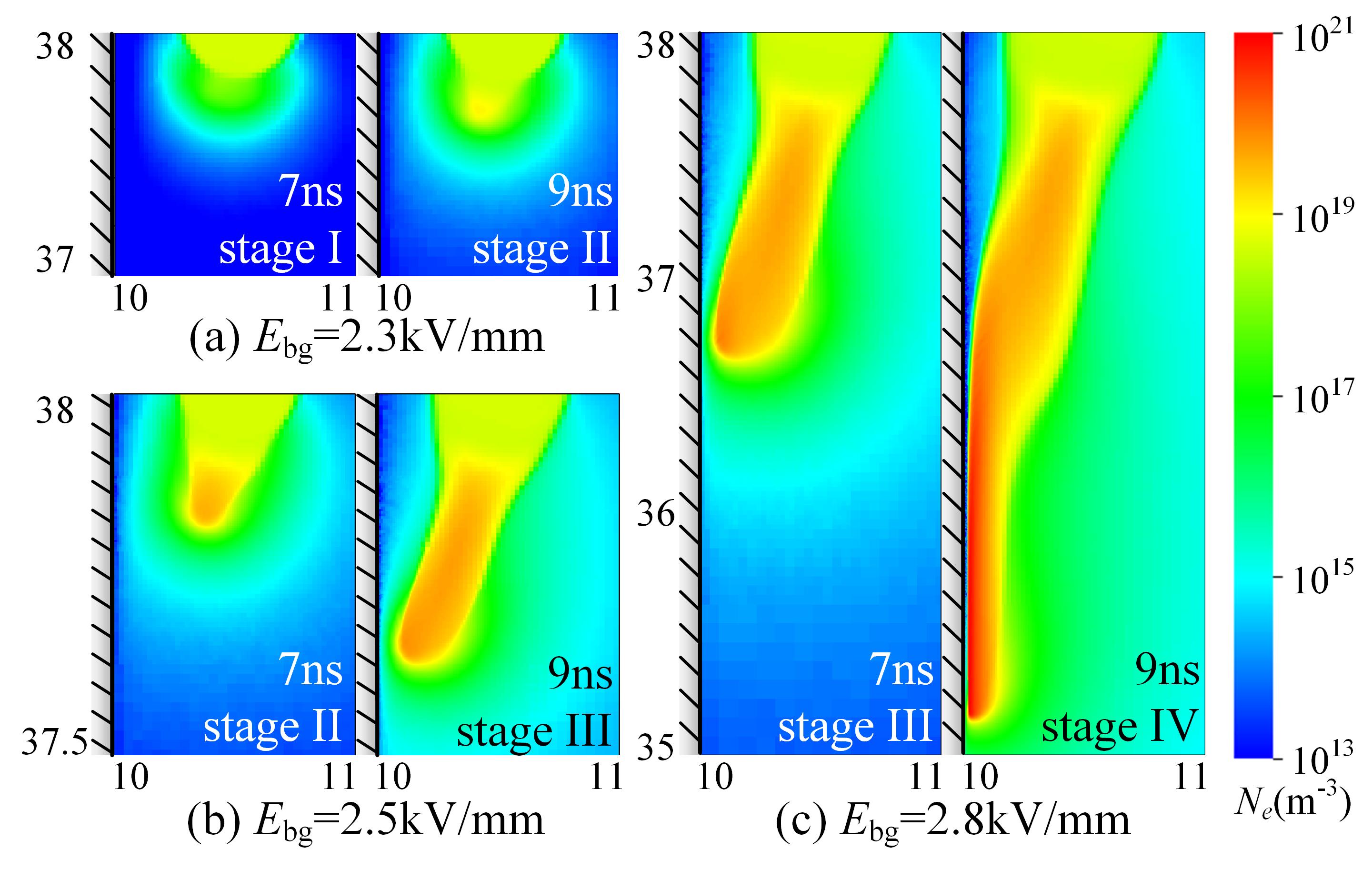}
	\caption{The electron density for streamers in different background electric fields ($2.3 \, \mathrm{kV/mm}$, $2.5 \, \mathrm{kV/mm}$ and $2.8 \, \mathrm{kV/mm}$) at $7 \, \mathrm{ns}$ and $9 \, \mathrm{ns}$.}
	\label{fig:2D_diffE}
\end{figure}

To study the effect of the applied voltage on surface discharges,
we have performed simulations for applied voltages of $92 \, \mathrm{kV}$, $100 \, \mathrm{kV}$ and $112 \, \mathrm{kV}$, which
correspond to background electric fields of $2.3 \, \mathrm{kV/mm}$, $2.5 \, \mathrm{kV/mm}$ and $2.8 \, \mathrm{kV/mm}$, respectively.
In all cases, the initial seed was located at $0.5 \, \mathrm{mm}$ from the dielectric, and the evolution up to $15 \, \mathrm{ns}$ was simulated.
Figure \ref{fig:maxE_t_diffE} shows the maximum electric field versus time, and \xiaoran{figure \ref{fig:2D_diffE} shows the electron density for three cases at $7 \, \mathrm{ns}$ and $9 \, \mathrm{ns}$. Both figures reveal the following stages in the streamer's development:}
\begin{enumerate}[I]
  \item The inception stage, in which the maximum electric field is from 0 to
  about $9 \, \mathrm{kV/mm}$ in our setup and the streamer is hardly propagating,  \xiaoran{as shown in figure \ref{fig:2D_diffE}a.}
  \item The gas-propagation stage, in which streamers propagate in the gas with
  a maximum electric field below $12.5 \, \mathrm{kV/mm}$.
  \xiaoran{This stage is visible in figures \ref{fig:2D_diffE}a and  figure \ref{fig:2D_diffE}b.
  }
  \item The transition stage from a gas streamer to a surface streamer, in which
  the maximum electric field increases sharply. The streamer also loses its
  rounded head shape, \xiaoran{as shown in figures \ref{fig:2D_diffE}b and \ref{fig:2D_diffE}c.}
  \item The surface propagation stage. The growth of the maximum electric field
  is slowing down, and the streamer propagates along the dielectric, as shown in figure \xiaoran{\ref{fig:2D_diffE}c.}
\end{enumerate}

Figure \ref{fig:maxE_t_diffE} shows that when the voltage is changed, the
streamers still exhibit similar behavior in these four stages. The main
difference is that the inception stage becomes shorter. For background fields of
$2.3 \, \mathrm{kV/mm}$, $2.5 \, \mathrm{kV/mm}$ and $2.8 \, \mathrm{kV/mm}$, the inception stages last $8.5 \, \mathrm{ns}$, $5.5 \, \mathrm{ns}$ and
$3.75 \, \mathrm{ns}$, respectively. The second stage also becomes slightly shorter for a
higher applied voltage.

\begin{figure}
  \centering
  \includegraphics[width=1.0\linewidth]{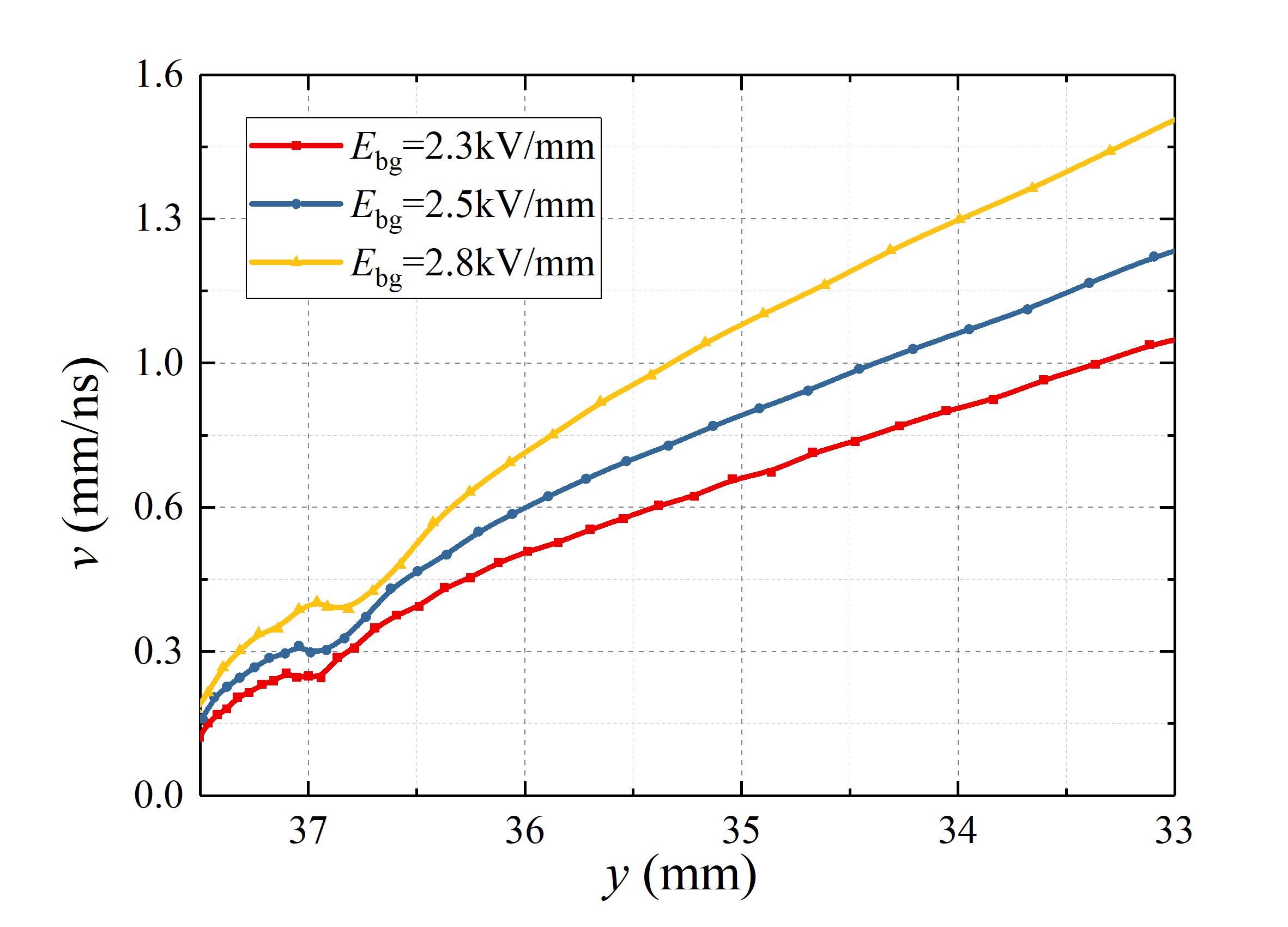}
  \caption{Streamer velocities versus their vertical location for different
    background electric fields.}
  \label{fig:v_y_diffE}
\end{figure}

\xiaoran{Figure \ref{fig:v_y_diffE} shows the streamer velocities versus their vertical location for the three applied voltages.
  As expected, a higher background electric field leads to a higher streamer velocity for streamers of the same length,
  in agreement with the experimental results of \cite{meng2015}.}

\subsection{Effect of dielectric permittivity}
\label{sec:effect-perm-diel}

The relative permittivity $\varepsilon$ of dielectric materials varies over a wide range.
To study how $\varepsilon$ affects surface discharges, we have performed
simulations with $\varepsilon$ set to 2, 3 and 5. As before, the initial
seed was placed at $0.5 \, \mathrm{mm}$ from the dielectric, and simulations were performed up
to $15 \, \mathrm{ns}$.
\begin{figure}
	\centering
	\includegraphics[width=1.0\linewidth]{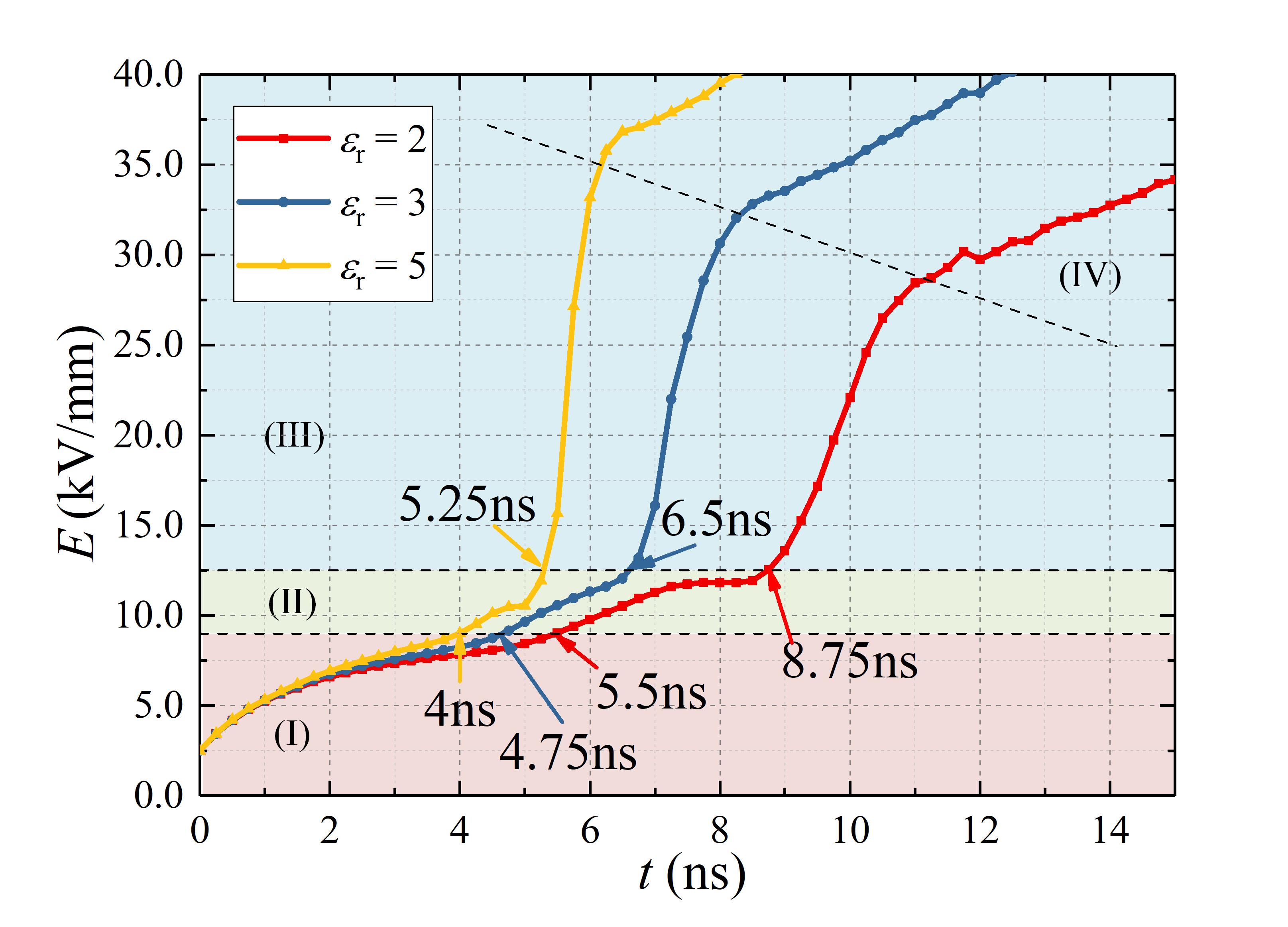}
	\caption{The streamers' maximal electric fields versus time for
          dielectrics with relative permittivities $\varepsilon_r$ of 2, 3 and
          5.}
	\label{fig:maxE-t-diffe}
\end{figure}

The maximum electric field versus time for the three permittivities is shown in
figure \ref{fig:maxE-t-diffe}. The main difference we observe is that the second
and third stages are shorter for a higher permittivity. A higher
$\varepsilon$ means the dielectric polarizes more strongly, which
leads to a stronger attraction of streamers to the dielectric.
Streamers therefore start the surface propagation stage earlier, and their maximum electric field increases more rapidly. Note that their maximum electric field is also higher during the surface propagation stage.
In summary, we can conclude that a higher permittivity leads to a faster transition into a surface streamer, and a higher maximum field during the surface propagation stage.

\begin{figure}
	\centering
	\includegraphics[width=1.0\linewidth]{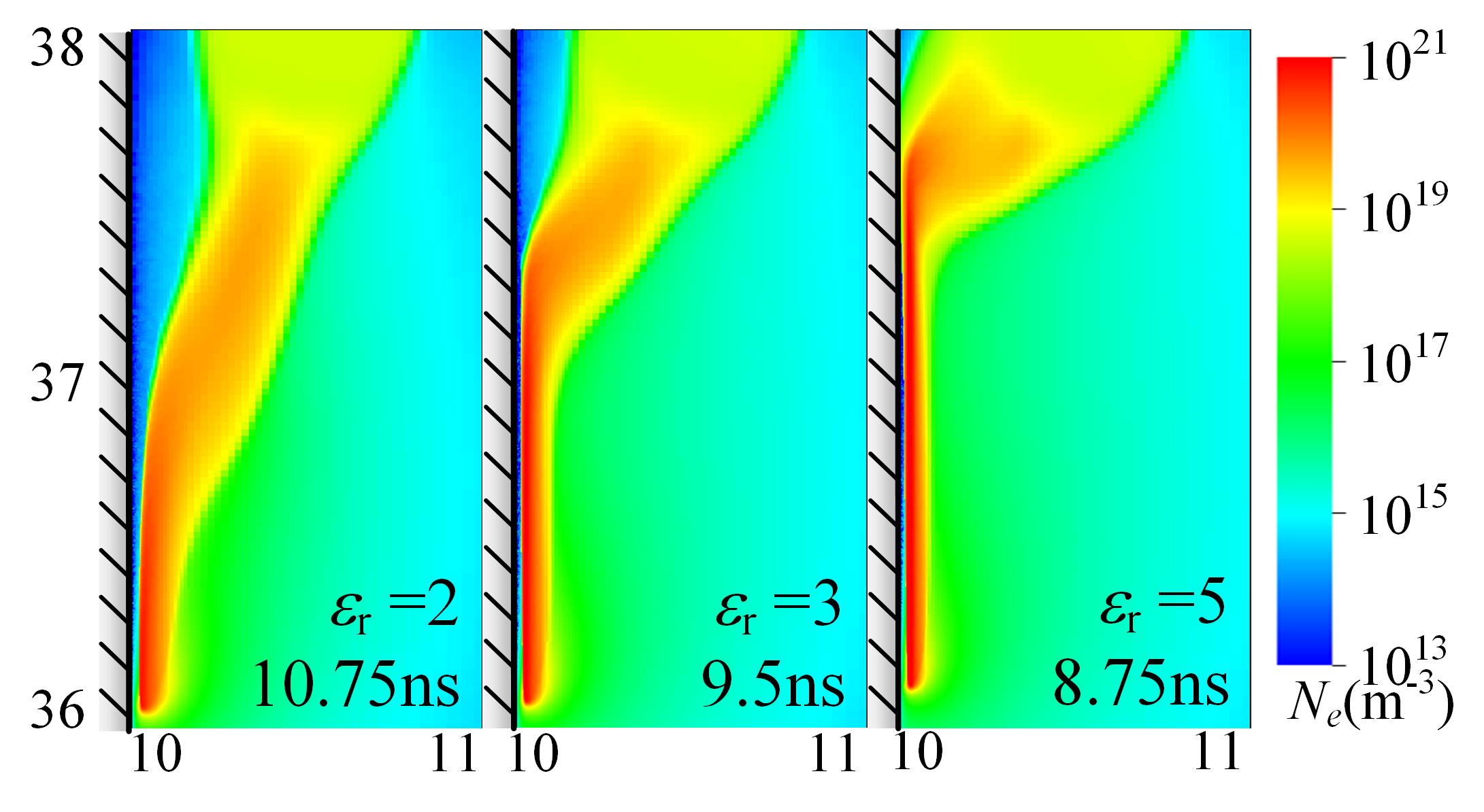}
	\caption{The streamers' electron density for dielectrics with relative
          permittivities $\varepsilon_r$ of 2, 3 and 5. Results are shown at
          different times, at the moment when the streamer heads are at
          $y = 36 \, \textrm{mm}$.}
	\label{fig:2D_diffeps}
\end{figure}

\xiaoran{Figure \ref{fig:2D_diffeps} shows the electron density for these three cases when all the streamers are at $y = 36 \, \mathrm{mm}$. It can clearly be seen that the streamers attach more rapidly to the dielectric with a higher permittivity. Notice also that the surface streamer's radius is smaller wither a higher dielectric permittivity, a result of the stronger electrostatic attraction.}

\begin{figure}
  \centering
  \includegraphics[width=1.0\linewidth]{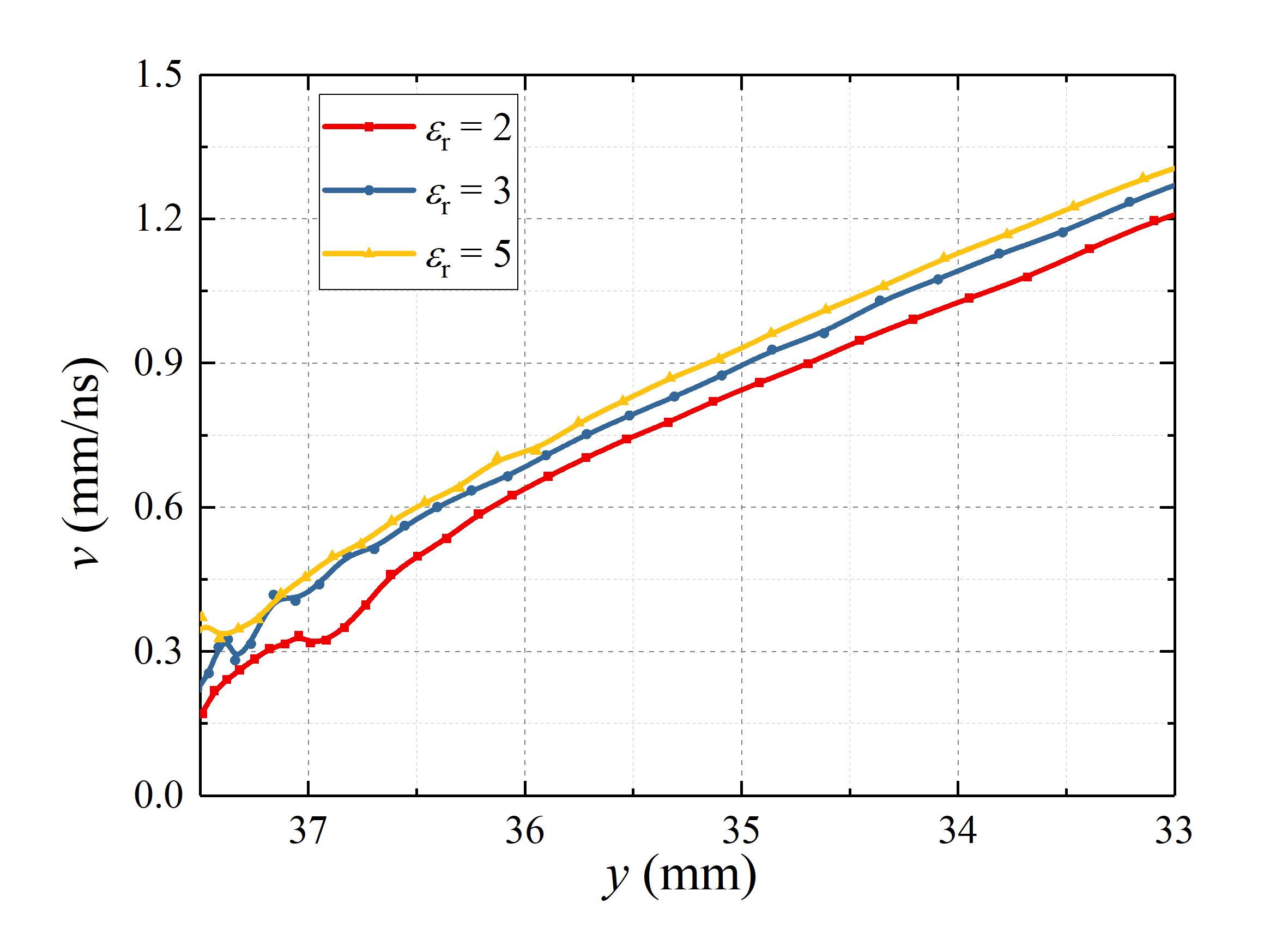}
  \caption{Streamer velocities versus their vertical location for different
    relative dielectric permittivities.}
  \label{fig:v_y_diffeps}
\end{figure}

\xiaoran{Figure \ref{fig:v_y_diffeps} shows the velocity versus the
  streamer's vertical position for the different $\varepsilon_r$. The
  permittivity has only a small effect on the streamer's velocity, in agreement
  with the experimental observations of~\cite{sobota2009}.
  In contrast, a negative correlation between the permittivity and the streamer velocity was found in~\cite{meng2015}. The discrepancy could come from the different geometry that was used, in which multiple surface and gas streamers propagated next to a cylindrical dielectric.}

\subsection{Effect of Secondary Electron emission from Dielectrics}
\label{sec:effect-electron-emission}

Electron emission from dielectrics may influence streamer velocities~\cite{tan2007} and
affect the high electric field in the dielectric-plasma gap~\cite{campanell2016}.
In this section, we study how secondary electron emission affects surface streamers in our computational geometry.
Both ion-induced secondary emission (ISEE) and photo-emission are considered.

\subsubsection{Ion-induced secondary electron emission}
\label{sec:electron-emission-ions}

\begin{figure}
	\centering
	\includegraphics[width=1.0\linewidth]{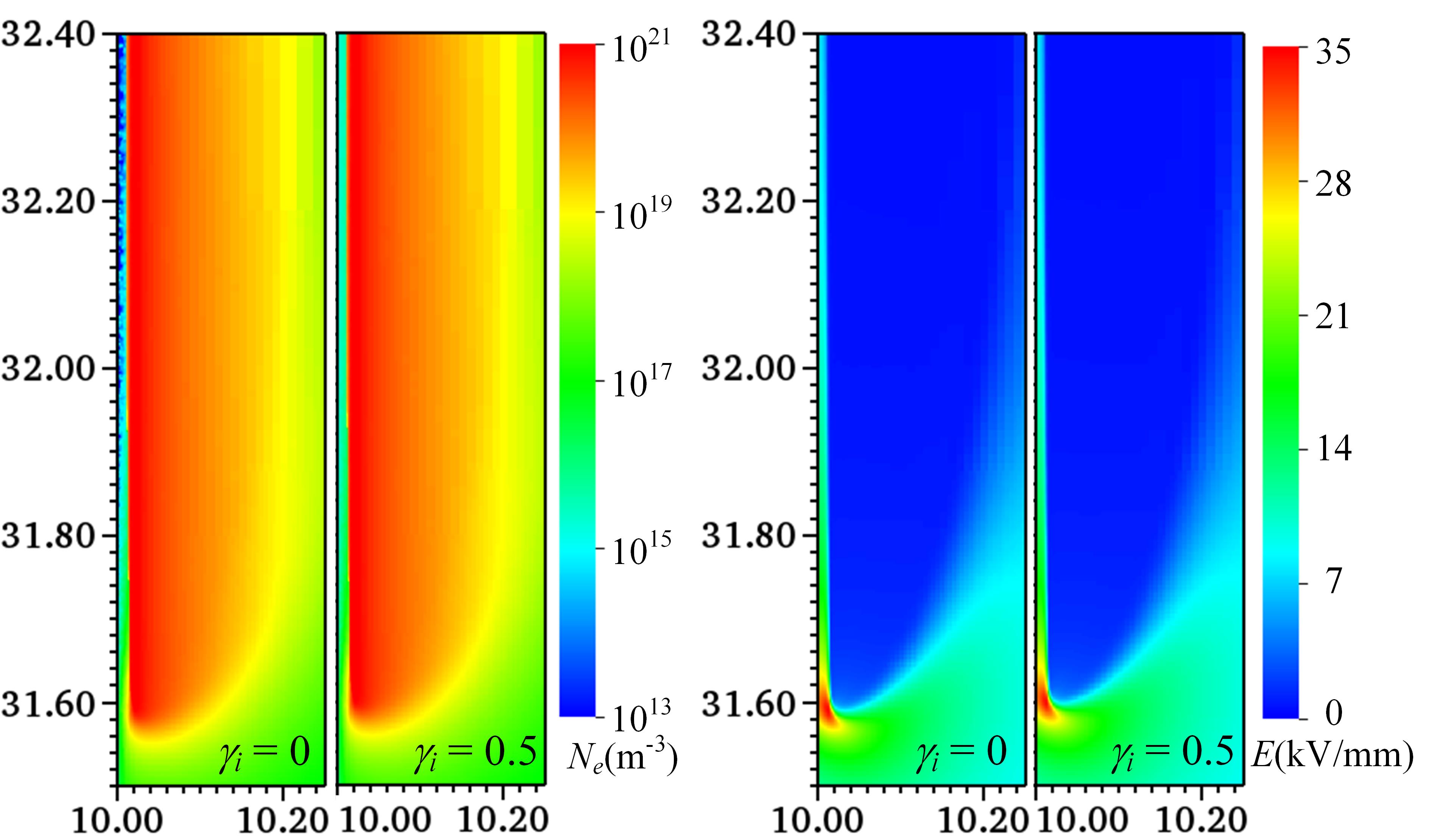}
	\caption{The electron density and electric field at $15 \, \mathrm{ns}$ with the
          ion-induced secondary electron emission coefficient $\gamma_{i}$ set
          to 0 (left) and $0.5$ (right).}
	\label{fig:isse}
\end{figure}

The ion-induced secondary electron emission (ISEE) yield $\gamma_{i}$ can vary over a wide range~\cite{motoyama2006, motoyama2004, tschiersch2017}.
Here we consider two cases, $\gamma_{i} = 0.5$ and $\gamma_{i} = 0$ (i.e., no secondary emission).
\xiaoran{In this section, the initial ionized seed's center is placed $0.4 \, \mathrm{mm}$ away from the dielectric, so that streamers start directly next to the dielectric.}
Figure \ref{fig:isse} shows the electron density and electric field distribution at $15 \, \mathrm{ns}$ for both ISEE coefficients.
It can be seen that ISEE here has little effect on the streamer length and the electric field.
The electron density in the streamer-dielectric gap is slightly higher behind the head for the streamer with $\gamma_{i} = 0.5$,
but this has negligible influence on the electric field in the gap.
\RV{We can conclude that the ISEE yield hardly affects the streamer head and its propagation (the latter is mainly determined by the former). The reason for this is that positive ions have a much lower mobility than electrons. Most positive ions are generated close to the streamer head, and they will not reach the dielectric in time to affect the rapidly propagating streamer head. Most ISEE electrons are therefore released after the streamer has passed by.}

\subsubsection{Photoemission}
\label{sec:electron-emission-photons}

The photo-emission coefficient ${\gamma}_{pe}$ for typical dielectric materials
varies between 10$^{-4}$ to 10$^{-1}$ for photon energies of 5--20 eV~\cite{fujihira1972,buzulutskov1997}.
This yield can be higher if the material contains stains or defects,
or when it is negatively charged~\cite{jorgenson2003,motoyama2006}.
The measurement of ${\gamma}_{pe}$ of dielectrics in air is often quite challenging~\cite{dubinova2016a}.
We here use several values for ${\gamma}_{pe}$ to demonstrate how photoemission affects positive streamers.

As discussed in section \ref{sec:photoionization-emission}, we consider low-energy and high-energy photons, with the main distinction that high-energy photons can be absorbed in the gas. The following four cases are considered for the photoemission coefficients ${\gamma}_{peL}$ and ${\gamma}_{peH}$ for low-energy and high-energy photons:
\begin{enumerate}
  \item ${\gamma}_{peH}$=0, ${\gamma}_{peL}$=0
  \item ${\gamma}_{peH}$=0.5, ${\gamma}_{peL}$=0
  \item ${\gamma}_{peH}$=0, ${\gamma}_{peL}$=0.5
  \item ${\gamma}_{peH}$=0.5, ${\gamma}_{peL}$=0.5
\end{enumerate}
In the simulations, streamers start near the dielectric, with the seed placed at either $0.5 \, \mathrm{mm}$ or $1 \, \mathrm{mm}$ from the dielectric.

\begin{figure*}
	\centering
	\includegraphics[width=0.9\linewidth]{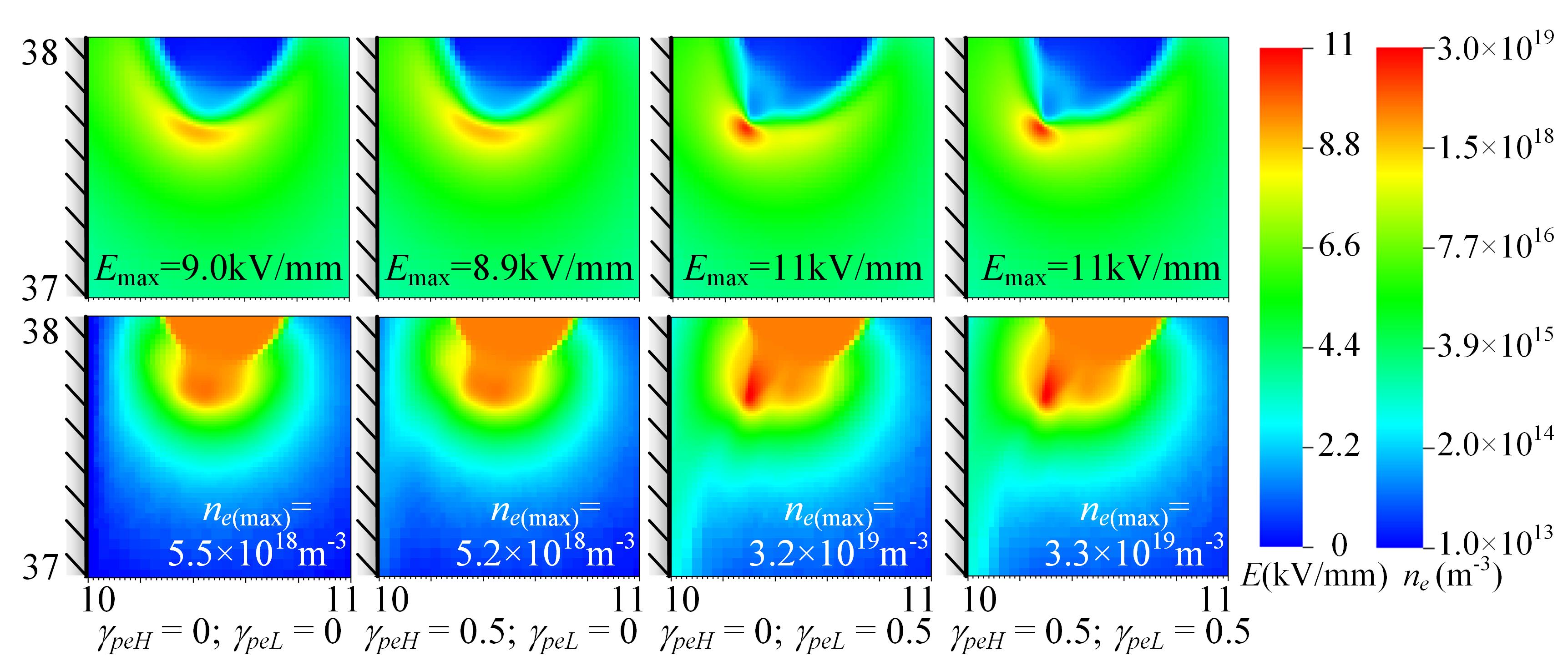}
	\caption{\RV{The electric field (top) and the electron density (bottom) in
          simulations with different photoemission coefficients for high-energy
          (${\gamma}_{peH}$) and low-energy photons (${\gamma}_{peL}$). Results
          are shown at $5.5 \, \mathrm{ns}$, when the streamer discharges start to grow.}}
	\label{fig:diff_phse_inception}
\end{figure*}

\begin{figure}
	\centering
	\includegraphics[width=1\linewidth]{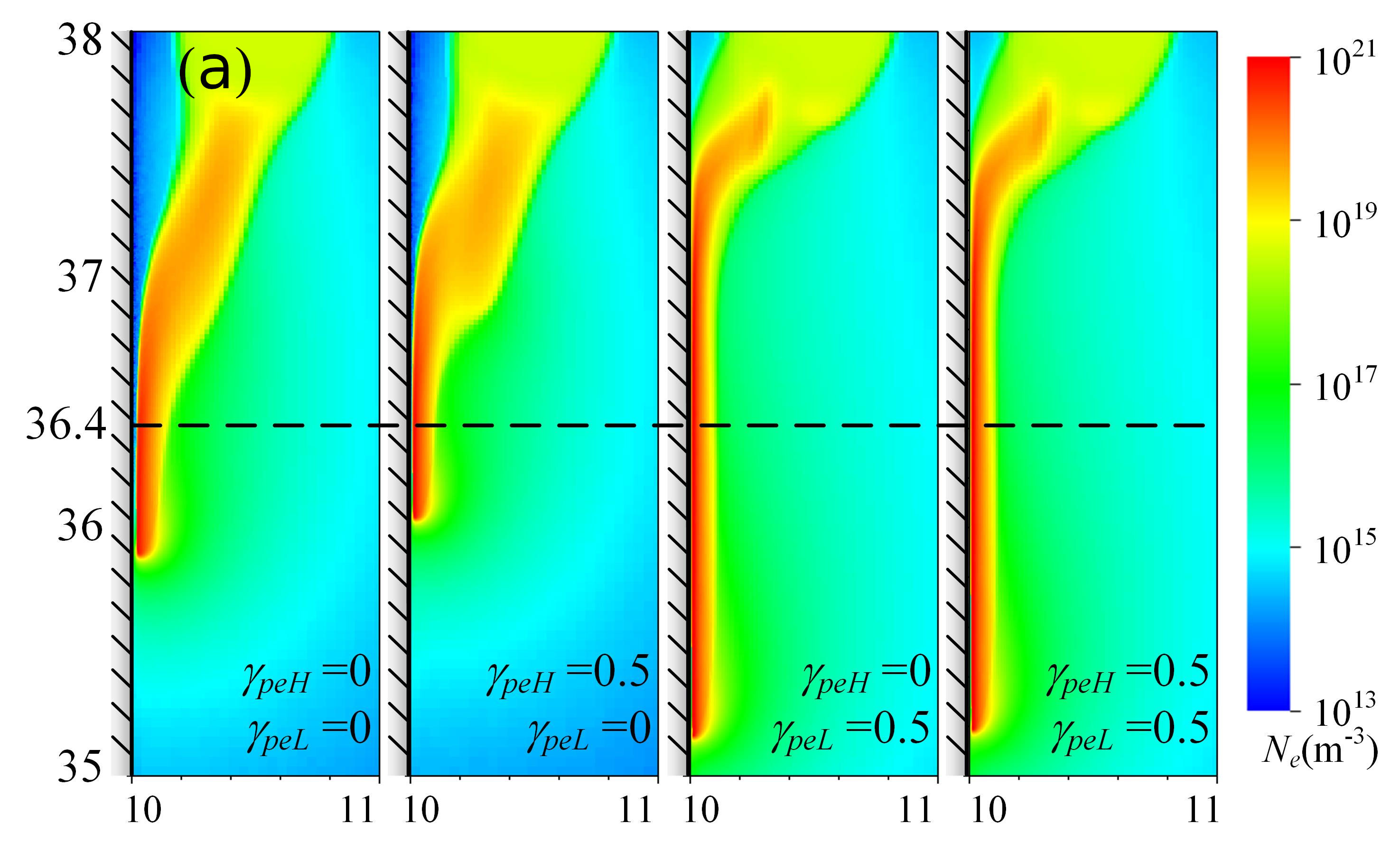}
        \includegraphics[width=1\linewidth]{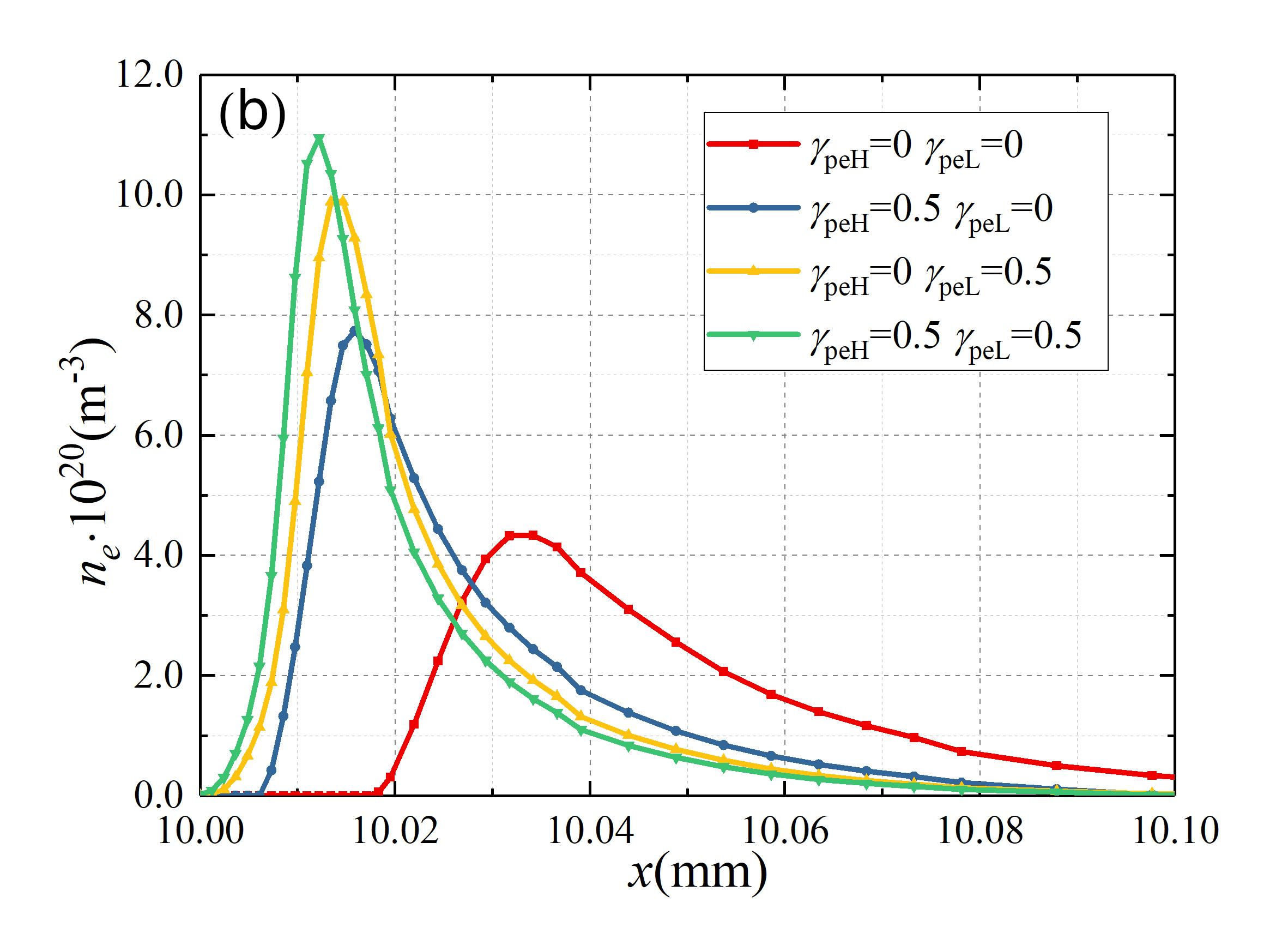}
        \caption{a) The electron density for simulations with different
          photoemission coefficients at $11 \, \mathrm{ns}$. b) Zoom of the electron density at
          $y = 36.4 \, \textrm{mm}$.}
	\label{fig:diff_phse}
\end{figure}

As shown in figure \ref{fig:diff_phse_inception}, photoemission by low-energy photons helps to start a discharge near a dielectric.
\xiaoran{At $5.5 \, \mathrm{ns}$, the ${\gamma}_{peL} = 0.5$ cases show the streamer already bending towards the dielectric with a sharp tip, due to photoemission.}
However, we remark that the inception time for these four cases is the same when the seeds are placed at $1 \, \mathrm{mm}$ away from the dielectric. The secondary electrons from the dielectric then need more time to reach the streamer, and the streamers have already started due to the photoionization they generate.
We conclude that photoemission can be important for discharges close to dielectrics and for discharges in gases with less photoionization than air.



\xiaoran{Figure \ref{fig:diff_phse}a shows the electron density distribution for the above four cases at $11 \, \mathrm{ns}$.
	The streamers with ${\gamma}_{peL} = 0.5$ are longer than the other two, since they start earlier. Photoemission also causes them to attach to the dielectric more rapidly.
Another difference is that the narrow gap between streamer and dielectric is smaller with more photoemission.
This happens because photoemission provides seed electrons in the gap, which allows the streamer to get closer to the dielectric.
To see this more clearly, the electron density distributions at $y = 36.4 \, \mathrm{mm}$ (the dashed line in figure \ref{fig:diff_phse}a) are shown for these four cases in figure \ref{fig:diff_phse}b.
Without photoemission, the electron density has a wider profile with a lower maximum, and it is located farther from the dielectric. When photoemission is included, the effect of the  low-energy photons (i.e., ${\gamma}_{peL} = 0.5$) is most important here.}

\begin{figure}
	\centering
	\includegraphics[width=1.0\linewidth]{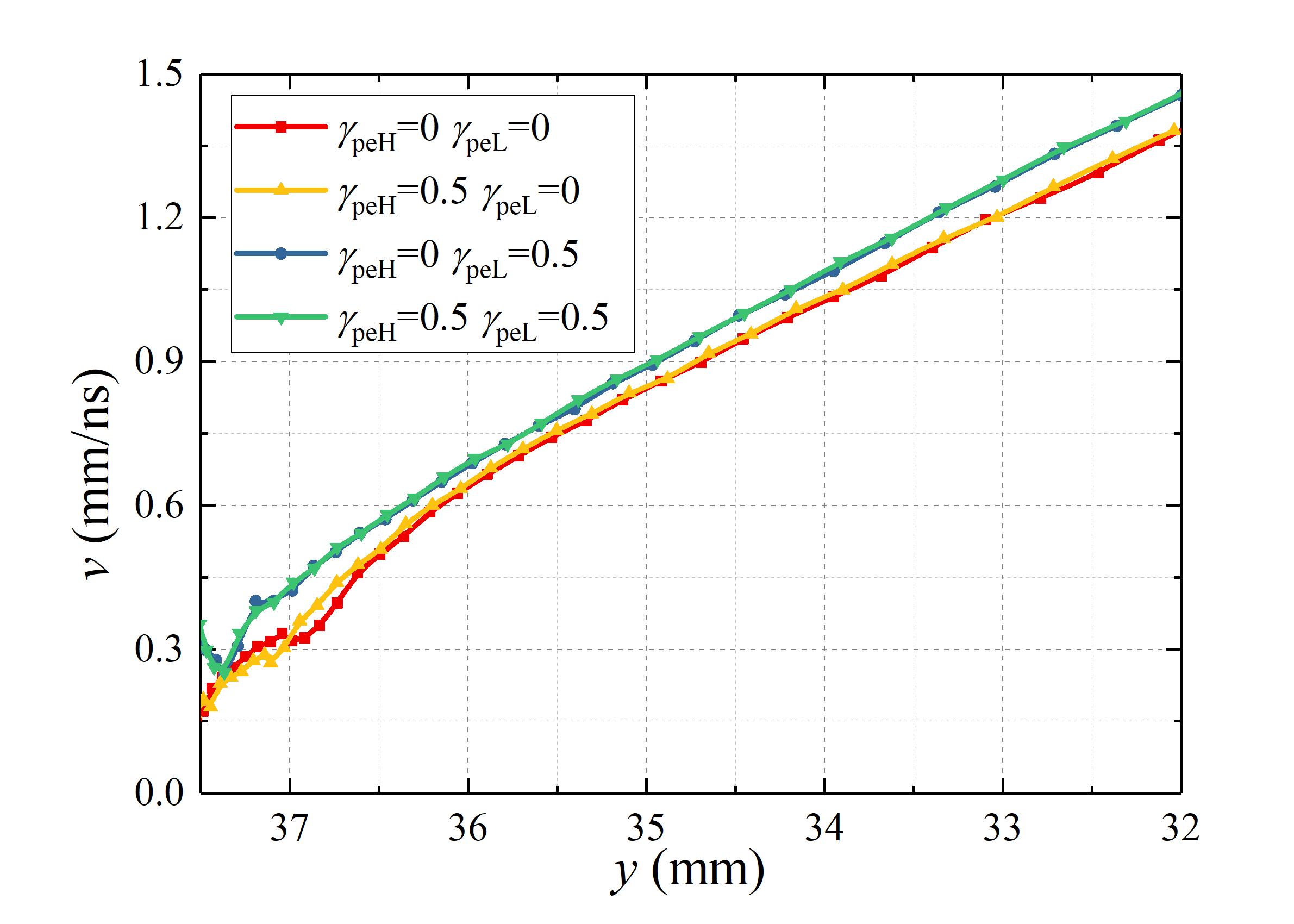}
	\caption{Streamer velocities versus their vertical location for different photoemission coefficients.}
	\label{fig:v_y_diffphse}
\end{figure}

\xiaoran{Figure \ref{fig:v_y_diffphse} shows the streamer velocities versus their vertical position for all four cases. The photoemission coefficients have only a small effect on the velocity, which is a little higher with $\gamma_{peL} = 0.5$.
  We think this is somewhat surprising. A possible explanation is that photoemission mostly leads to growth towards the dielectric, whereas photoionization in the gas contributes most of the free electrons that cause growth parallel to the dielectric.
  Another effect in the simulations presented here is that high-energy photons contribute less to a streamer's growth very close to a dielectric. There are two reasons for this. First, these photons are absorbed at shorter distances if they hit a dielectric. Second, their photoemission coefficient is here less than one ($\gamma_{peH} = 0.5$), whereas in the gas they always lead to photoionization.
}
\subsection{Effect of positive ion mobility}
\label{sec:effect-positive-ion}

\begin{figure}
  \centering
  \includegraphics[width=0.8\linewidth]{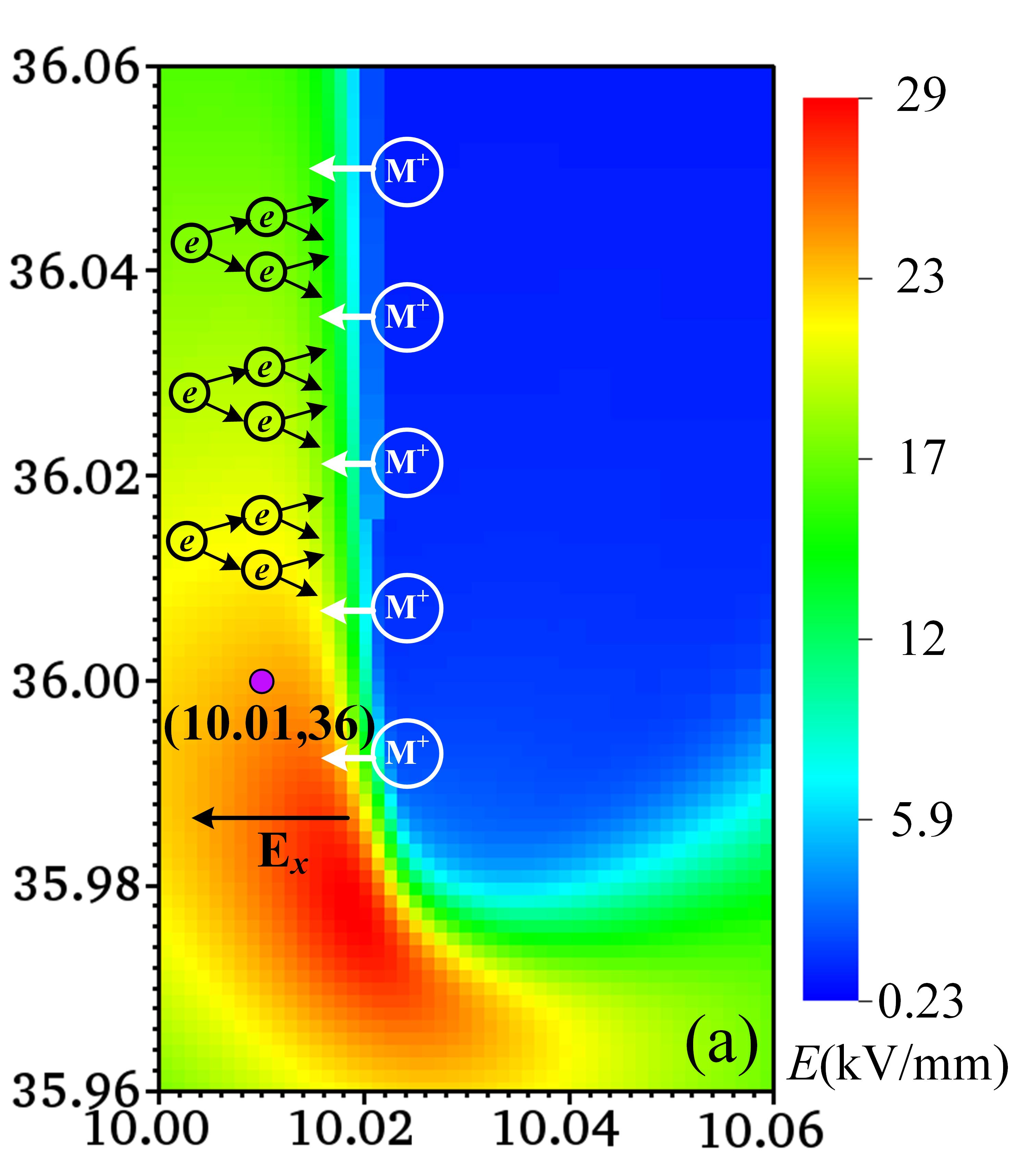}
  \includegraphics[width=1.0\linewidth]{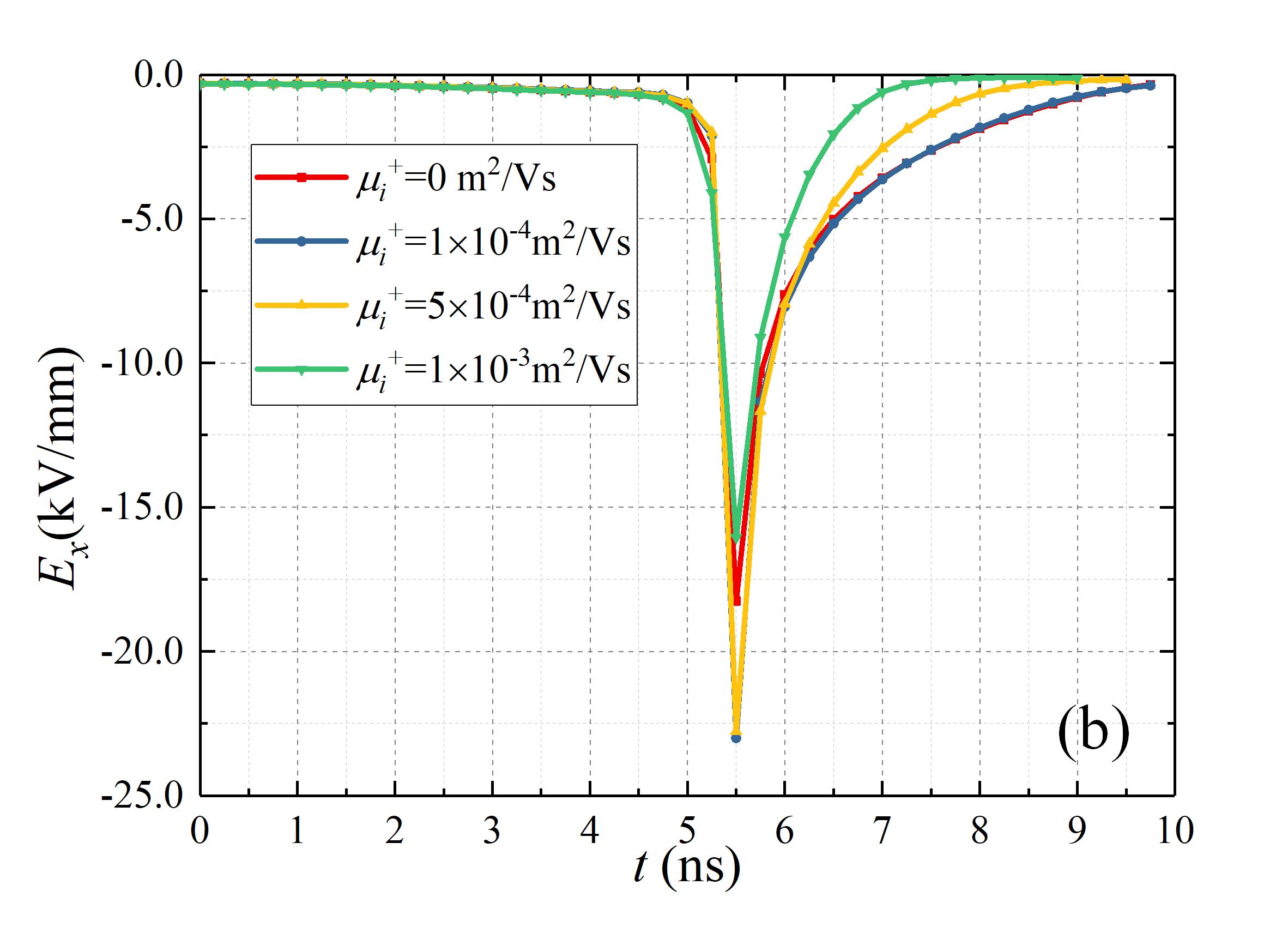}
  \caption{a) The electric field at $5.5 \, \mathrm{ns}$ after streamer inception for a
    positive ion mobility of 3${\times}$10$^{-4 }$m$^{2}$/Vs. The electron and
    positive ion dynamics in the streamer-dielectric gap are illustrated. b) The
    $E_{x}$ field at the point
    $(x, y) = (10.01 \, \mathrm{mm}, 36 \, \mathrm{mm})$ versus time for
    streamers with different positive ion mobilities. Here $t=0$ corresponds to
    the streamers' respective inception times, which vary by less than a
    nanosecond for the four cases. }
  \label{fig:ion_mobility_gap}
\end{figure}

The positive ion mobility $\mu_i^+$ can affect surface streamers in two ways. First, a higher ion mobility increases the amount of ion-induced secondary electron emission (ISEE). However, since ISEE was found to play a negligible role in section \ref{sec:electron-emission-ions}, its role is not further studied here, and we set the ISEE yield to zero (i.e., $\gamma_i = 0$).

A second effect is that a higher ion mobility increases the conductivity of the discharge, in particular in regions where the ion density is high compared to the electron density. For positive surface streamer discharges, such a region is present in the streamer-dielectric gap. This gap typically contains a high electric field, especially close to the streamer head, see section \ref{sec:cathode-sheath}.
Electrons rapidly drift away from the surface, whereas positive ions move from the high-density discharge region towards the surface, as illustrated in figure \ref{fig:ion_mobility_gap}a.

To investigate how the positive ion mobility ($\mu_i^+$) affects the decay of the high electric field in the streamer-dielectric gap,
we have performed simulations with positive ion mobilities of 0, 1${\times}$10$^{-4 }$m$^{2}$/Vs, 5${\times}$10$^{-4 }$m$^{2}$/Vs and 1${\times}$10$^{-3}$ m$^{2}$/Vs, using a seed placed $0.5 \, \mathrm{mm}$ from the dielectric.
For these simulations, we have recorded $E_{x}$ in the middle of the gap at the point indicated in figure \ref{fig:ion_mobility_gap}a. The recorded fields are shown versus time in figure \ref{fig:ion_mobility_gap}b.
The maximum electric field occurs when the streamer heads pass by the observation point indicated in figure \ref{fig:ion_mobility_gap}a.
The decay of the peak in $E_{x}$ is faster for higher $\mu_i^+$, which is most clearly visible for the $\mu_i^+ = 5 \times 10^{-4} \, \mathrm{m}^{2}/\mathrm{Vs}$ and $\mu_i^+ = 1 \times 10^{-3} \, \mathrm{m}^{2}/\mathrm{Vs}$ cases.
Note that the field also decays when the ions are immobile. This mainly happens because the amount of net space charge is lower behind the streamer head, but electron avalanches in the gap (due to e.g. photoionization) also contribute.

\section{Conclusions}
\label{sec:conclusions}

In this paper, we have studied positive surface streamers with numerical simulations.
A 2D fluid model for surface discharges based on the Afivo-streamer code ~\cite{teunissen2017} was developed. The model includes a Monte Carlo procedure for secondary electron emission (from both high and low energy photons) and support for dielectric surfaces. These new features are compatible with the adaptive mesh refinement and the parallel multigrid solver provided by the underlying Afivo framework~\cite{teunissen2018}.

We have used the new model to investigate the interaction \RV{of} positive streamers and dielectrics. We considered a parallel-plate geometry, with a flat dielectric between the two electrodes. Positive streamer discharges started from an ionized seed that was placed near the dielectric and the positive electrode.
The effect of several parameters was investigated: the applied voltage, the dielectric permittivity, secondary electron emission caused by ions and photons, and the mobility of positive ions.
Our main findings are summarized below:
\begin{enumerate}
  \item A narrow gap forms between positive streamers and dielectrics, as was also observed in earlier work~\cite{babaeva2016,stepanyan2014,soloviev2017}. A very high electric field can be present in this so-called `cathode sheath'.
  \item The attraction of positive streamers to the dielectric was found to be mostly electrostatic. In our geometry, this attraction was caused by the net charge in the streamer head, which polarized the dielectric, increasing the field between the streamer and the dielectric. A higher dielectric permittivity led to a more rapid attachment of the streamer to the dielectric.
  \item Compared to gas streamers, surface streamers had a smaller radius, a higher electric field, and a higher electron density. In our simulations, this gave surface streamers a higher propagation velocity than gas streamers.
  \item A higher applied voltage caused the positive surface discharges to start earlier, but they behaved qualitatively similar. A higher dielectric permittivity also accelerated the formation of surface streamers.
  \item Photoemission can accelerate streamer inception near dielectrics. However, photoemission hardly increases the velocity of surface streamers. A possible reasons is that photoemission mostly leads to growth towards the surface, whereas photoionization contributes more to the growth parallel to the surface.
  \item The positive ion mobility affects the decay of the high electric field in the streamer-dielectric gap.
\end{enumerate}

\section*{Acknowledgment}

This project was supported by \RV{the National Natural Science Foundation of China (51777164), the State Key Laboratory of Electrical Insulation and Power
Equipment (EIPE18203)} and the Fundamental Research Funds for the Central Universities of China (xtr042019009). \RV{JT was also supported by fellowship 12Q6117N from
Research Foundation -- Flanders (FWO).}

\section*{Availability of model and data}

The source code and input files for the model used in this paper are available at \url{https://gitlab.com/MD-CWI-NL/afivo-streamer} (git commit \texttt{08f1c828}).

\section*{References}

\bibliography{references}

\end{document}